\documentclass[a4paper]{article}
\usepackage{amsmath}
\usepackage{graphicx}

\usepackage{hyperref}
\usepackage{bm}

\newcommand*{\affaddr}[1]{#1} % No op here. Customize it for different styles.
\newcommand*{\affmark}[1][*]{\textsuperscript{#1}}
\newcommand*{\email}[1]{\texttt{#1}}

\newcommand{\appropto}{\mathrel{\vcenter{
  \offinterlineskip\halign{\hfil$##$\cr %The $##$ is not a mistake!
    \propto\cr\noalign{\kern2pt}\sim\cr\noalign{\kern-2pt}}}}}

\newcommand{\ssim}{\,{\sim}\,} % special sim %Gaps reduced, looks much better.
\begin{document}
% Don't want date printed
%\date{}
% Make title large and bold
%\title{\Large\bfseries Paper Name Here}
\title{Ice Shelves as Floating Channel Flows of Viscous Power-Law Fluids}

% Target typesetting:
%
% Author A, Author B, Author C, Author D and Author E
%        A,B,C Department of Computer Science
%       D,E Department of Mechanical Engineering
%          Email A,B,C,D,E @university.edu
%                  Latex University
%\maketitle
%\bgroup\setlength{\parindent}{0pt}
\author{Indranil Banik\affmark[1] \& Justas Dauparas\affmark[1]\\
\affaddr{\affmark[1] Institute of Theoretical Geophysics, Department of Applied Mathematics\\ and Theoretical Physics, Wilberforce Road, Cambridge CB3 0WA, UK} \\
\email{Email:  \href{mailto:ib45@st-andrews.ac.uk}{ib45@st-andrews.ac.uk} (IB)},~~~~~ \email{\href{mailto:jd540@cam.ac.uk}{jd540@cam.ac.uk} (JD)}}
%\email{\{A,B,C,D,E\}@university.edu}\\}
\maketitle

%\maketitle

%\author{Indranil Banik$^{1}$\footnote[*]{Email: \href{mailto:ib45@st-andrews.ac.uk}{ib45@st-andrews.ac.uk} (Indranil Banik)\newline $~~~~~~~~~~~~~~$ \href{mailto:hz4@st-andrews.ac.uk}{jd540@cam.ac.uk} (Justas Dauparas)} \& Justas Dauparas$^{1}$\\
%$^{1}$Scottish Universities Physics Alliance, University of St Andrews, North Haugh, St Andrews, Fife, KY16 9SS, UK}

%$^{1}$Scottish Universities Physics Alliance, University of St Andrews, North Haugh, St Andrews, Fife, KY16 9SS, UK
\tableofcontents

\newpage
\clearpage

\begin{abstract}

We explain the force balance in flowing marine ice sheets and the ice shelves they often feed. Treating ice as a viscous shear-thinning power law fluid, we develop an asymptotic (late-time) theory in two cases $-$ the presence or absence of contact with sidewalls. Most real-world situations fall somewhere between the two extreme cases considered. The solution when sidewalls are absent is a fairly simple generalisation of that found by Robison (JFM, 648, 363). In this case, we obtain the equilibrium grounding line thickness using a simple computer model and have an analytic approximation. For shelves in contact with sidewalls, we obtain an asymptotic theory valid for long shelves. We determine when this is. Our theory is based on the velocity profile across the channel being a generalised version of Poiseuille flow, which works when lateral shear dominates the force balance.

We conducted experiments using a laboratory model for ice. This was a suspension of xanthan in water, at a concentration of $0.5\%$ by mass. The model has $n \approx 3.8$, similar to that of ice. Our theories agreed extremely well with our experiments for all relevant parameters (front position, thickness profile, lateral velocity profile, longitudinal velocity gradient and grounding line thickness). We also saw detailed features similar to natural systems. Thus, we believe we have understood the dominant force balance in both types of ice shelf.

Combining our understanding of the forces in the system with a basic model for basal melting and iceberg formation, we uncovered some instabilities of the natural system. Laterally confined ice shelves can rapidly disintegrate but ice tongues can not. However, ice tongues can be shortened until they no longer exist, at which point the sheet becomes unstable and ultimately the grounding line should retreat above sea level. While the ice tongue still exists, the flow of ice into it should not be speeded up and the grounding line should also not retreat, assuming that only conditions in the ocean change. However, laterally confined ice shelves experience significant buttressing. If removed, this leads to a rapid speedup of the sheet and a new equilibrium grounding line thickness. We believe that something like this occurred in the Larsen B ice shelf.

%Our work has implications for the collapse of ice shelves, such as Larsen B. Lateral friction in the shelf provides significant buttressing, which can be removed if the shelf collapses. This would greatly accelerate the flow of ice and accelerate sea level rise. Our model allows a calculation of the magnitude of the effect based on the topography and the initial rate of flow, if such a collapse were to occur. However, for an ice tongue, sudden acceleration of this nature appears unlikely as there is no lateral friction. Problems can still arise if changing conditions shorten the ice tongue until it no longer exists, because then the formation of icebergs at the front of the ice sheet leads to an instability and loss of all ice below sea level.

\end{abstract}

\newpage
\clearpage
\section{Introduction}

This paper builds on previous work by Robison \cite{Robison_2010} on what are essentially ice tongues i.e. ice shelves with no lateral confinement. We generalise this to fluids with arbitrary shear-thinning coefficient $n$ (our theory should also work for shear-thickening fluids, but we did not perform experiments using such fluids). We then generalise the asymptotic result for laterally confined ice shelves found by Pegler \cite{Pegler_2013}. The ice sheet is treated as a classical viscous gravity current, though we need to be careful about when this is a valid assumption.

In this paper, we start with a first principles theoretical solution for shelves without sidewalls in the case of constant initial thickness (Section \ref{Ice_tongues}). We also provide some reasons for expecting the initial thickness to be constant. Then, we derive a similarity solution for a laterally confined shelf in a channel of constant width (Section \ref{Laterally_confined_ice_shelves}). The solution is valid in the asymptotic limit, so we also derive roughly where this is (Equation \ref{L_sidewall}).

The sheet is briefly reviewed so that we can address the grounding line (Section \ref{Grounding_line}). The equilibrium grounding line thickness is found for ice tongues (Equation \ref{G}). For laterally confined ice shelves, we already have the initial thickness and thus the grounding line position without considering the sheet. We briefly discuss how the sheet may influence the dynamics, noting that it does not in the asymptotic limit (Section \ref{Grounding_line_sidewalls}).

\begin{figure} [b]
	\centering
		\includegraphics [width = 16cm] {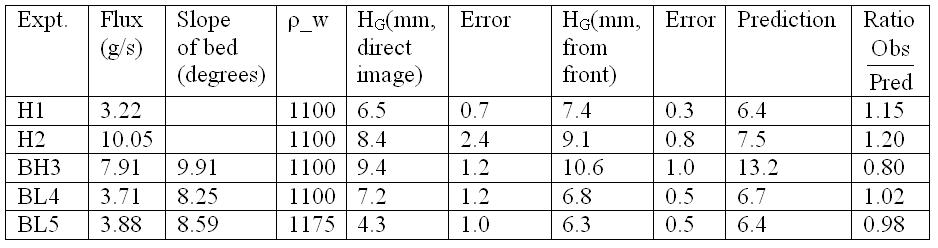}
	\caption{Results obtained from experiments without lateral friction. We only have a very basic understanding of experiments without a sloped bed (the first two listed above). Lateral thickness variations make our model work poorly for BH3. Note the minor impact of the change in $g'$ (buoyancy) between the last two experiments (xanthan at the concentration used has $\rho = 995$ kg/$m^3$).}
	\label{Ice_tongue_data}
\end{figure}

We describe $\ssim 50$ experiments we conducted to help us develop these theories and to test them (Section \ref{Experiments}). The experiments with a $1\%$ mass concentration xanthan are described first, for the case of sidewall contact (Section \ref{One_percent_xanthan}). Then, the effect of halving the concentration is shown (Section \ref{Half_percent_xanthan}). Data for experiments in which there was no sidewall contact are also shown and compared with theoretical predictions for the grounding line thickness (Figure \ref{Ice_tongue_data}). An important point is that there are artefacts of our experimental setup due to the way the flow is initialised. We determine the length over which such effects are dissipated (Equation \ref{L_weir}). Fortunately, the sheet was longer than this length.

\begin{figure}[ht]
	\centering
		\includegraphics [width = 10cm] {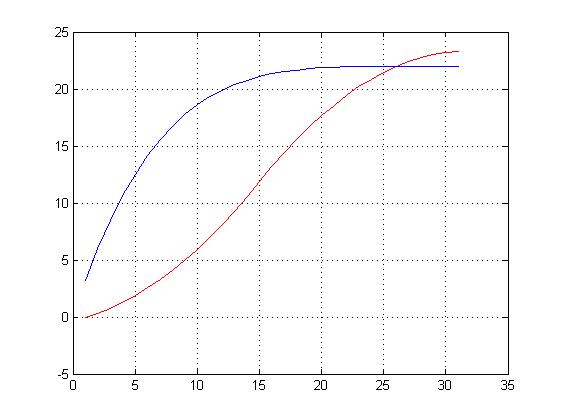}
		\includegraphics [width = 10cm] {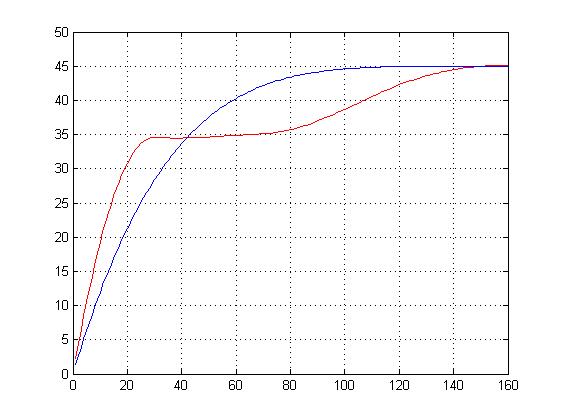}	
		\caption{Vertical velocity profiles $u \left( z \right)$ obtained in the sheet are shown in red, near the weir (\emph{top}) and further away (\emph{bottom}). The component of the velocity along the channel is shown. In blue, we try to fit the expected velocity profile for a viscous gravity current, by matching velocities at the top of the sheet. Note that the data are from an experiment with sidewall contact (this should not be important, as the sheet should be vertical shear dominated).}
	\label{Sheet_Profile}
\end{figure}

As well as the position of the propagating front, we also show thickness of the shelf as a function of position (the profile). This is based on a photograph which agrees rather well with our model (Section \ref{Thickness_Profile}). To help confirm our model, we used particle imaging velocimetry (PIV) to determine the velocity field in the shelf (Figure \ref{Velocity_Profile}) and sheet (Figure \ref{Sheet_Profile}). The methods used to obtain this data and the results are discussed and compared with theoretical expectations in Section \ref{PIV}.

%As well as the position of the propagating front, we also show the velocity field in the shelf (Figure \ref{Velocity_Profile}) and sheet (Figure \ref{Sheet_Profile}). The methods used to obtain this data and the results are discussed and compared with theoretical expectations (Section \ref{PIV}). The thickness of the shelf as a function of position (the profile) is also shown from a photograph and compared with our model (Figure \ref{Real_Profile}).

Using our newly developed understanding, we give a partial explanation of how a collapsing ice shelf affects the rate of flow of the associated ice sheet (Section \ref{Towards_nature}). This should be treated with some caution at this stage, but can readily explain large increases in the flow rate over short time intervals (such as occurred with Larsen B). Our work suggests that such a phenomenon can only occur in ice shelves significantly affected by sidewall contact, something which is easily checked. Our model allows a rough calculation of the magnitude of the effect if a particular ice shelf were to collapse, based on topography and other data. However, our work does not shed much light on which shelves are likely to actually collapse.

\newpage
\clearpage
\section{Ice Tongues}
\label{Ice_tongues}

%The figure should be the first thing, and must fit on the first page at all costs.
\begin{figure} [ht]
%\centering
  \includegraphics[width = 16.5cm]{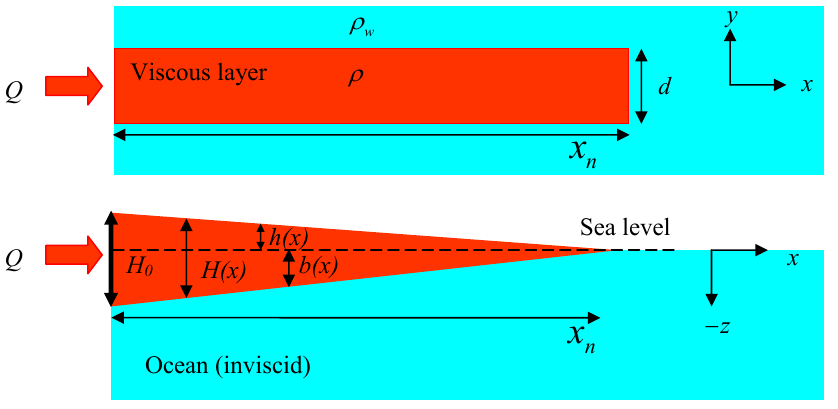}
\caption{Plan and side views of the situation considered.}
\label{Primary_Situation}
\end{figure}

We consider the flow of an incompressible viscous fluid at low Reynolds number according to the geometry shown in \text{Figure} \ref{Primary_Situation}. We take the flux $Q$ to be held constant and assume that the flow does not spread laterally. Although the width of the flow probably could be determined based on $Q$ and other parameters, we do not attempt to do this. Instead, we treat the width of the shelf $d$ as an independent variable. 

Using $\bm{r} \equiv \left(x,y,z \right)$ for position and $\bm{v} \equiv \left(u,v,w \right)$ for velocity, our viscosity model is
\begin{eqnarray}
	\eta ~=~ \eta_o \left[\sqrt{\frac{1}{8} \sum_{i=x,y,z}~\sum_{j=x,y,z}\left( \frac{\partial \bm{v}_i}{\partial \bm{r}_j} + \frac{\partial \bm{v}_j}{\partial \bm{r}_i} \right)^2}~\right]^{\frac{1}{n}-1}
	\label{Viscosity_law}
\end{eqnarray}

This is motivated by laboratory studies of ice indicating that it can be modelled quite well in this way using $n = 3.2 \pm 0.1$, with temperature changes mainly serving to change $\eta_o$ rather than $n$ \cite{Glen_1955}.

The force balance along the channel ($x$-direction) is that 
\begin{eqnarray}
	\frac{\partial }{\partial x}\int_{-b}^{h}{{{\sigma }_{xx}}~dz}~=~-{{\rho }_{w}}gb\frac{\partial b}{\partial x}
\end{eqnarray} 

because there are no lateral or vertical stresses and water pressure has a component in the $x$-direction (as the normal to the shelf does). We neglect lateral flow and assume that $H' \equiv \frac{\partial H}{\partial x} \ll 1$ so that $w \ll u$ (vertical flow negligible). Because the value of the above integral at the front of the shelf must balance with the integrated hydrostatic pressure of the ocean \cite{Robison_2010}, we have that
\begin{eqnarray}
  \int_{-b}^{h}{{\sigma }_{xx}}~dz&=&-\frac{1}{2}{{\rho }_{w}}g{b}^{2}=-\frac{1}{2}\rho gHb, ~\text{       because } \rho_{w}b = \rho H \text{ (Archimedes' Principle)} 
\end{eqnarray}

Considering the absence of lateral and vertical shear in this system, Equation \ref{Viscosity_law} simplifies to
\begin{eqnarray}
	\eta ~=~ \eta_o { \left(\frac{\partial u}{\partial x} \right)}^{\frac{1}{n}-1}
	%\label{Viscosity_law}
\end{eqnarray}

Writing that
\begin{eqnarray}
	{{\sigma }_{xx}} ~\equiv ~ -P+2\eta \frac{\partial u}{\partial x}
\end{eqnarray} 

and applying a vertical balance of forces argument to ${{\sigma }_{zz}} ~\equiv ~ -P+2\eta \frac{\partial w}{\partial z}$ (using also $\frac{\partial w}{\partial z} = -\frac{\partial u}{\partial x}$ due to continuity) we obtain that 
\begin{eqnarray}
	{{\sigma }_{xx}}~=~-\rho g(h-z) ~+~ 4\eta \frac{\partial u}{\partial x}
\end{eqnarray} 

Integrating this vertically, we get that 
\begin{eqnarray}
 -\frac{1}{2}\rho g{{H}^{2}}+\int 4\eta \frac{\partial u}{\partial x}~dz&=&-\frac{1}{2}\rho gHb \\ 
 4{{\eta }_{o}}{{\int{\left( \frac{\partial u}{\partial x} \right)}}^{\frac{1}{n}}}~dz&=&~\frac{1}{2}\rho gH(H-b) \\ 
  &=&~\frac{1}{2}\rho g'{{H}^{2}},~\textrm{ because }h\equiv\frac{\text{ } g'}{g}H
	\label{Equation_8}
\end{eqnarray}

The term $g'$ is called the reduced gravity and accounts for the fact that only a fraction of the shelf is above the waterline. Thus, gradients in the height above sea level are smaller than gradients in $H$. Applying Archimedes' Principle to the shelf, we get that
\begin{eqnarray}
	\frac{g'}{g} ~=~ \frac{{\rho}_{w}-\rho}{{\rho}_{w}}
\end{eqnarray}

Continuing with our derivation,
\begin{eqnarray}
  {{\eta }_{o}}{{\left( \frac{\partial u}{\partial x} \right)}^{\frac{1}{n}}}H~&=&~\frac{\rho g'{{H}^{2}}}{8} \\ 
	  \frac{\partial u}{\partial x}~&=&~{{\left( \frac{\rho g' H}{8{{\eta }_{o}}} \right)}^{n}} \label{A}
\end{eqnarray}

As this is positive, we note that $u > 0$ throughout the shelf. We now use this information inside the continuity equation.
\begin{eqnarray}
	{d}^{-1} \frac{\partial H}{\partial t}\ +~ H\frac{\partial u}{\partial x}\ +~ u\frac{\partial H}{\partial x} ~=~ 0 
\end{eqnarray}

We use a Lagrangian picture to better visualise the situation.	
\begin{eqnarray}
{d}^{-1} \frac{DH}{Dt}\ &=& -H\frac{\partial u}{\partial x}\ \\
 &=& -{{H}^{n+1}}\alpha, ~\textrm{ where }\alpha \equiv {{\left( \frac{\rho g'}{8{{\eta }_{o}}} \right)}^{n} \textrm{  is constant.}}
\end{eqnarray}

In the co-moving (Lagrangian) frame, each fluid element enters the shelf at time ${{t}_{0}}$. Assuming that H$({{t}_{0}})$ is independent of $t$ (i.e. a constant source thickness), we must have that $H = H(t-{{t}_{0}})$ only. Using Equation \ref{A}, we see that $\left.\frac{\partial u}{\partial x} \right|_{t}=f(t-{{t}_{0}})$. Therefore,
\begin{eqnarray}
  u(x,t) &=& {{u}_{0}} + \int_{0}^{x}{f(t-{{t}_{0}})~dx'} \\ 
 &=& {{u}_{0}} + \int_{0}^{x}{f(\tau (x'))~dx'}\text{   where }\tau \equiv t-{{t}_{0}}  \text{  and   }\tau = 0   \text{   for    } x=0
\end{eqnarray}

We now convert the integral required to obtain $u$ from one over $x$ to one over $\tau$. The value of $\tau$ is 0 when $x = 0$. When the fluid element reaches $x$, $\tau \equiv t - {t}_{0}$. We note that a fluid element injected at the source over a time interval $d\tau$ has a total volume $Q~d\tau$. At all later times, it occupies the same volume. However, it is also a part of the profile. This means that it occupies a volume $Hd~dx$. Thus,
\begin{eqnarray}
 u({{t}_{0}},t) &=& {{u}_{0}} + \int_{0}^{t-{{t}_{0}}}{f(\tau )\frac{dx'}{d\tau }\ d\tau } \\
 &=& {{u}_{0}} + \int_{0}^{t-{{t}_{0}}}{f(\tau )\frac{Q}{H(\tau)~d}\ d\tau }
\end{eqnarray}

This means that $u$ is a function of ($t-t_0$) only, as we know that $H$ is and $u_{0}=\frac{Q}{{{H}_{0}}d}$ is assumed constant. Integrating $u$ and assuming that the source of the shelf (a grounding line) remains static, we see that $x$ is also going to be a function of $\ t-{{t}_{0}}$ only. All fluid elements reach a given $x$ at the same value of $\tau$. Thus, at that position, $H$ is always the same (after the front has reached this position). We therefore have a steady profile. Under these conditions, the continuity equation reduces to
\begin{eqnarray}
	u ~=~ \frac{Q}{Hd}
\end{eqnarray}

Differentiating this with respect to $x$ and substituting in Equation \ref{A}, we obtain a first-order differential equation for the profile. Solving this subject to the constant initial thickness $H_0$, we get that 
%We emphasise that this profile has a typical length scale $L$.
\begin{eqnarray}
	H &=& {{\left[ \frac{Q}{(n+1)\alpha\ d} \right]}^{\frac{1}{n+1}}}{{\left( x+L \right)}^{-\frac{1}{n+1}}}\label{Ice_tongue_profile}\\
	u &=& {{\left( \frac{Q}{d} \right)}^{\frac{n}{n+1}}}{{\left[ \left( n+1 \right)\alpha d \right]}^{^{\frac{1}{n+1}}}}{{\left( x+L \right)}^{\frac{1}{n+1}}} 
\end{eqnarray}

The constants $L$ and $\alpha$ are defined below:
\begin{eqnarray}
	L ~&=&~ \frac{Q}{(n+1)\alpha d{{H}_{0}}^{n+1}}\\
	\alpha ~&\equiv&~ {\left( \frac{\rho g'}{8{{\eta }_{o}}} \right)}^{n}
\end{eqnarray}

When $x_n \ll L$, the profile is essentially flat and the speed equals the initial value ($\frac{Q}{{H}_{0}d}$). For $x_n \gg L$, we have that
\begin{eqnarray}
	\label{eq:1}
	H &\sim& {{x}^{-\frac{1}{n+1}}} \\
	u &\sim& {x}^{\frac{1}{n+1}}
\end{eqnarray}

Finally, the position of the front as a function of time is readily determined from the velocity field.
\begin{eqnarray}
	{{x}_{n}} ~=~ \frac{Q}{(n+1)\alpha\ H_0^{n+1}d}\left[ {{\left( 1+\alpha n{{H}_{0}}^{n}t \right)}^{^{\frac{1}{n}+1}}}-1 \right]
\end{eqnarray}

Alternatively, we could ensure the area enclosed by the profile upstream of the front is correct. As the profile is steady, this entails solving
\begin{eqnarray}
	\int_{0}^{{{x}_{n}}(t)}{H(x)dx ~=~ \frac{Qt}{d}}
\end{eqnarray}

Of course, both approaches agree for all $n$. For the case of $n = 1$, our solution reduces to that found by \cite{Robison_2010}.

In a real system, the shelf would be fed by a sheet at a grounding line (the `source'). The thickness here completely determines the buttressing exerted on the sheet and also affects the velocity field in the sheet. There is likely to be a unique thickness at which the forces at the grounding line are in equilibrium. Once this is attained, there is no further tendency for change (as the buttressing is independent of how far the front has propagated $-$ it is always $\frac{1}{2}\rho g' {{H}_{0}}^{2}$). We assume that the equilibrium so attained is stable. The equilibrium grounding line thickness is calculated in Section 4.1, although perturbations are not considered in this work.

%Basically, if the grounding line is too deep, the pushing force is too low and water pressure is too high, forcing the grounding line back upstream. Not sure how to encode this mathematically.

\newpage
\clearpage
\section{Laterally Confined Ice Shelves}
\label{Laterally_confined_ice_shelves}

The geometry in this situation is the same as that considered before, except that now the half-width of the shelf rather than the full width is $d$. The major difference is the presence of laterally confining sidewalls, which we assume the shelf is in contact with at all times.

The effect of sidewalls will dominate over the effects of hydrostatic pressure on an ice shelf if the shelf is long enough relative to its width and height. To get an estimate for when this may be the case, we determine when the shelf starts thickening. In order for this to happen, the velocity field must be altered such that instead of $\frac{\partial u}{\partial x}>0$, we have $\frac{\partial u}{\partial x}<0$, at least temporarily. This means that, rather than continuity forcing the shelf to thin with distance, the front is going slower than fluid elements behind it so that the flow essentially ‘piles up’. We expect this to occur in order to provide a pressure gradient to overcome the viscous drag from sidewalls and keep the shelf flowing.

The key thing is that ${{\sigma }_{xx}}$ is not purely hydrostatic pressure. There is a difference between pressure from xanthan and that from water (partly due to their different densities). This is balanced with a non-zero value of $\frac{\partial u}{\partial x}$. As we have seen, this is (initially) positive.

Now, if we were to reduce ${{\sigma }_{xx}}$ enough that $\frac{\partial u}{\partial x}$ were forced to become negative, then the situation would indeed be different to the no sidewalls scenario. To achieve this, a certain amount of drag from sidewalls is required. Note that $\frac{\partial {{\sigma }_{xx}}}{\partial x}+\frac{\partial {{\sigma }_{xy}}}{\partial y}=0$. For $y>0$, $\frac{\partial u}{\partial y}<0$ so ${{\sigma }_{xy}}<0$. Noting that the surface $y=0$ is free by symmetry, we see that $\frac{\partial {{\sigma }_{xy}}}{\partial y}<0$, this also holding for $y<0$. Thus, $\frac{\partial {{\sigma }_{xx}}}{\partial x}>0$. Assuming $H$ is not yet altered (so neither is $\frac{\partial u}{\partial x}$ at the front), this means that $\frac{\partial u}{\partial x}$ at the source eventually goes negative and becomes increasingly so. Once this occurs, the front starts decelerating and the shelf will be forced to start thickening, invalidating our assumption of a flat shelf.

As before, the initial value of that part of ${{\sigma }_{xx}}$ caused by the $\frac{\partial u}{\partial x}$ term is $\frac{1}{2}\rho g'{{H}^{2}}$ when integrated vertically (Equation \ref{Equation_8}). Thus, for the shelf to start thickening, the total force from sidewalls must exceed the lateral integral of the above term (the total non-hydrostatic force, or pushing force). This way, there will no longer be any pushing force at all. With an even longer shelf and even more drag, it will change sign, making $\frac{\partial u}{\partial x}<0$. We assume that this is a good indicator of when sidewalls start to have a significant impact upon the dynamics of the flow.
\begin{eqnarray}
	\frac{1}{2}\rho g'{{H}^{2}}.2d ~=~ {{\eta }_{o}}{{\left( \frac{1}{2}\frac{\partial u}{\partial y} \right)}^{\frac{1}{n}-1}}\left( \frac{\partial u}{\partial y} \right).2LH
\end{eqnarray}

Henceforth, if we raise a negative number to a power, we mean that the absolute value of the number is raised to that power. \emph{The end result is always positive}. For the viscosity, we have assumed that $u=0$ along the walls (the no slip condition) so only lateral variations contribute to the total strain there. As a rough estimate, we assume a triangular velocity profile.
\begin{eqnarray}
	\frac{\partial u}{\partial y} ~=~ \frac{2\overline{u}}{d}
\end{eqnarray}

This is an underestimate because the boundary layer is probably very thin so has more shear. Using also the fact that $\overline{u}=\frac{Q}{2Hd}$, we obtain that
\begin{eqnarray}
	L ~=~ \frac{\rho g'{{H}^{^{1+\frac{1}{n}}}}{{d}^{^{1+\frac{2}{n}}}}}{{{2}^{2-\frac{1}{n}}}{{\eta }_{o}}{{Q}^{^{\frac{1}{n}}}}}
	\label{L_sidewall}
\end{eqnarray}

We note that $\frac{\partial u}{\partial x}$ is very small in the asymptotic limit, because the shelf will get longer and longer while the start of the shelf gets thicker, so $u$ there decreases. As the force balance equation must always hold, we expect (treating ${{\sigma }_{xx}}$ as purely hydrostatic) that the above equation linking $L$ and some typical (e.g. maximum) thickness should hold in the asymptotic limit, when it becomes accurate tp assume that $\frac{\partial u}{\partial x}$ is irrelevant to the $\eta$ as it only adds in quadrature to the dominant $\frac{\partial u}{\partial y}$ term (Equation \ref{Viscosity_law}). Essentially, we have balanced the hydrostatic pressure discontinuity not with a $\frac{\partial u}{\partial x}$ term but instead with a $\frac{\partial u}{\partial y}$ term. This suggests that the system is self-similar at late times.

We have said that $x_n \gg L$ is required for sidewalls to dominate the shelf, but assuming that a solution is discovered valid for such situations, when does this solution first become an accurate description of the length of the shelf? One probably needs computer simulations to answer this in detail, but here we give a rough idea. The shelf can not be longer than the solution predicts, because then $\left|H'\right|$ is even lower so there is insufficient driving force to overcome viscous drag from the sidewalls. Also, even more drag exists than in the solution, because the shelf is even longer. So the situation can not arise. 

However, the shelf can be shorter than our solution predicts $-$ to compensate for the lower drag, it can simply be thick at the front, reducing the thickness gradient. So we see that there is no problem with a shelf shorter than the solution predicts, but a major problem with longer shelves $-$ these can't exist, if the force balance is dominated by sidewalls. Thus, one way to determine $L$ may be to find when continuing to apply the no sidewalls (basically, constant $u$) solution leads to the front being further ahead than the yet to be derived solution when sidewalls dominate. Because this situation is impossible, we can use this to determine the point at which the no sidewalls solution can no longer be applied to the system. This approach would require determining the initial thickness of the shelf, which is possible under some circumstances (see Section 4.1).

%If there is one message to take away, it is that theories don't always work, so knowing when they should (and when they shouldn't) is essential.

%Later on, say that we opted to measure H directly early in an experiment to calculate L for that experiment. Obviously, also give a table of the multiple of L by which convergence occurred (and errors). But if it is converging at 10L, then a slightly imperfect start causing loss of contact for 1cm will lead to convergence occurring 10cm later, so we can not expect all experiments to converge at precisely the same multiple of L. The initial thickness is not very interesting, as it is surface tension dominated, but say that we tried to get a partial understanding of the situation.

%The value of $L$ obtained from the above equation is consistent with experimental observations, which indicate a tendency for the shelf to start thickening and for the front to slow down when it is about $L$ from the source, and converge upon a self-similar mode of propagation within a few times this distance

We begin by using the fundamental equation for the balance of forces along the channel. 
\begin{eqnarray}
	\frac{\partial {{\sigma }_{xx}}}{\partial x} + \frac{\partial {{\sigma }_{xy}}}{\partial y} ~=~ 0
\end{eqnarray}

We assume that $\frac{\partial u}{\partial x}\ll\frac{\partial u}{\partial y}$ over the vast majority of the channel, because the shelf is much longer than it is wide. This allows us to consider only lateral stresses. Also assuming negligible transverse and vertical velocities and no lateral variations in thickness, we obtain that
\begin{eqnarray}
	\frac{\partial }{\partial y}\left[ {{\eta }_{o}}{{\left( \frac{1}{2}\frac{\partial u}{\partial y} \right)}^{\frac{1}{n}-1}}\frac{\partial u}{\partial y} \right]  ~=~ \rho g'\frac{\partial H}{\partial x} \label{BB}
\end{eqnarray} 

Notice that only gradients in $H$ can affect the velocity field, because the resulting pressure gradients \emph{alone} drive the flow. Integrating the above equation with respect to $y$ and applying the no-slip boundary condition for $y = \pm d$ as well as no lateral stress along the centreline of the channel due to symmetry (i.e. $\frac{\partial u}{\partial y}=0\text{ for }y=0$), we obtain the velocity profile:
\begin{eqnarray}
	u ~=~ \frac{{2}^{1-n}}{n+1}{{\left( \frac{\rho g'H'}{{{\eta }_{o}}} \right)}^{n}}\left( {{d}^{n+1}}-{{y}^{n+1}} \right) ~~\text{   where   } H'\equiv\frac{\partial H}{\partial x}
	\label{Equation_33}
\end{eqnarray}

We shall refer to a flow with a velocity pattern like this as a generalised Poiseuille flow, the classical example being when $n = 1$. The flux crossing a plane of constant $x$ is easily found to be
\begin{eqnarray}
	q(x)&=&2H\int_{0}^{d} u~dy\\
	&=&\frac{{2}^{2-n}}{n+2}{{\left( \frac{\rho g'}{{{\eta }_{o}}} \right)}^{n}} {d}^{n+2}\left( H{H'}^{n} \right) \label{Shelf_flux}
\end{eqnarray}

The continuity equation can now be applied to obtain a single non-linear diffusion equation for the fluid. 
\begin{eqnarray}
	\frac{\partial H}{\partial t} ~+~ \frac{{{2}^{1-n}}{{d}^{n+1}}}{n+2}{{\left( \frac{\rho g'}{{{\eta }_{o}}} \right)}^{n}}{{\left( H{H'}^{n} \right)}{'}} ~=~ 0
\end{eqnarray}

For a more complicated geometry, a computer simulation will be required to understand what happens, although future work on simple geometries and on slowly varying $d$ could shed some light on the problem. In such work, the ${d}^{n+1}$ term should be brought inside the last bracket to allow for the possibility that the width of the channel varies with position. For now, $d$ is constant. 

We now apply scaling arguments to the above equation and to the equation of global mass conservation
\begin{eqnarray}
	\int_{0}^{{{x}_{n}}(t)}{H(x)dx ~=~ \frac{Qt}{2d}}
\end{eqnarray}

This suggests that the following quantity is a dimensionless constant of order 1:
\begin{eqnarray}
	\frac{{{x}_{n}}{{(n+2)}^{\frac{1}{2n+1}}}{{2}^{\frac{2n-1}{2n+1}}}}{{{t}^{\frac{n+1}{2n+1}}}{{d}^{\frac{1}{2n+1}}}{{Q}^{\frac{n}{2n+1}}}}{{\left( \frac{{{\eta }_{o}}}{\rho g'} \right)}^{\frac{n}{2n+1}}}
\end{eqnarray}

We look for a solution in terms of the similarity variable
\begin{eqnarray}
	\varepsilon ~\equiv ~ \frac{x{{(n+2)}^{\frac{1}{2n+1}}}{{2}^{\frac{2n-1}{2n+1}}}}{{{t}^{\frac{n+1}{2n+1}}}{{d}^{\frac{1}{2n+1}}}{{Q}^{\frac{n}{2n+1}}}}{{\left( \frac{{{\eta }_{o}}}{\rho g'} \right)}^{\frac{n}{2n+1}}}
\end{eqnarray}

Our analysis indicates that this is directly proportional to $\frac{x}{{x}_{n}}$. Because ${x}_{n}$ rises slower than $t$, the fact that the area enclosed by the profile must rise linearly with time implies that the whole profile must also be thickening. Thus, $H$ will necessarily have an explicit dependence on $t$. Using the fact that $H\sim\frac{Qt}{2dx_{n}}$, we obtain that
\begin{eqnarray}
	H ~=~ \frac{{{(n+2)}^{\frac{1}{2n+1}}}{{Q}^{\frac{n+1}{2n+1}}}{{t}^{\frac{n}{2n+1}}}}{{{2}^{\frac{2}{2n+1}}}{{d}^{\frac{2n+2}{2n+1}}}}{{\left( \frac{{{\eta }_{o}}}{\rho g'} \right)}^{\frac{n}{2n+1}}}\psi (\varepsilon )
	\label{Equation_41}
\end{eqnarray}

where the dimensionless profile $\psi (\varepsilon )$ is of order 1 near the source and decreases to 0 at the front. Differentials in $x$ and $t$ can be converted into differentials in $\varepsilon$ by applying the usual chain rule. Such an analysis shows that the powers of all externally imposed parameters are indeed equal on all terms. This allows us to obtain a single ordinary differential equation for the profile in terms of similarity co-ordinates.
\begin{eqnarray}
  {{(\psi \psi {{'}^{^{n}}})}'}&=&-\frac{n}{2n+1}\psi (\varepsilon )+\frac{n+1}{2n+1}\varepsilon \psi '(\varepsilon )\label{C}\\ 
  \int_{0}^{{{\varepsilon }_{n}}}{\psi (\varepsilon )d\varepsilon }&=&1 \\
  \psi{{\psi'}^{n}}&=&1 \text{     at } \varepsilon = 0 \label{D}
\end{eqnarray}

The term $\psi \psi {{'}^{^{n}}}$ corresponds to the (dimensionless) flux crossing a given position. This gradually decreases from its initial value. The reason is that part of it is `lost along the way' because it goes into increasing the thickness of the profile.

The advance of the front is not driven by the requirement to push flux through it (unlike in ice tongues). Instead, it is driven by the fact that $H' \neq 0$ there, leading to a non-zero velocity of fluid elements at the front (Equation \ref{Equation_33}).

We obtain approximate expressions for $\psi'$ that become exact at either end of the profile. Near the source (or the rear) of the profile,
\begin{eqnarray}
	\psi ' ~\approx ~ -{{\left( \frac{1}{\psi \left( 0 \right)} \right)}^{\frac{1}{n}}}
\end{eqnarray}

Near the front (at ${{\varepsilon }_{n}}$), we may obtain a first integral of Equation \ref{C} to deduce that
\begin{eqnarray}
	\psi ' ~\approx ~ -{{\left( \frac{n+1}{2n+1}{{\varepsilon }} \right)}^{\frac{1}{n}}}
\end{eqnarray}

We have used the fact that the integral of $\psi$ with respect to $\varepsilon$ (from the front to a point nearby) is second order in the value of $\psi$, as the profile is approximately triangular in this region (a singularity in $H'$ leads to a singular velocity profile, so $H'$ must be finite). Using these results, we may obtain an expression for the total change in $\psi'$ over the profile. This yields the fractional change in velocity along the profile because $u\propto \psi {{'}^{^{n}}}$. We will need the actual value of ${\varepsilon}_{n}$ to compute this.

We solved Equation \ref{C} numerically by shooting backwards from the front, using as boundary conditions $\psi =0$ at $\varepsilon = {\varepsilon}_{n}$ and the above expression for $\psi '$. Due to computing errors, there is a small error in $\psi \psi {{'}^{^{n}}}$ at the source, but none at the front of the profile. Obviously, errors near the source are much preferred because $\psi \psi {{'}^{^{n}}} = 0$ at the front (so we can ill afford errors here).

Although the solution looks reasonable for almost any value of ${{\varepsilon }_{n}}$, only one value can actually make the total area enclosed by the profile equal to 1. An estimate for the error made by the computer can then be obtained by checking how far off the solution is from satisfying Equation \ref{D}.

%Indicate that the error is less than 10^-5 or something like this.

\newpage

\begin{figure}[ht]
  \centering
  \includegraphics[width = 10cm]{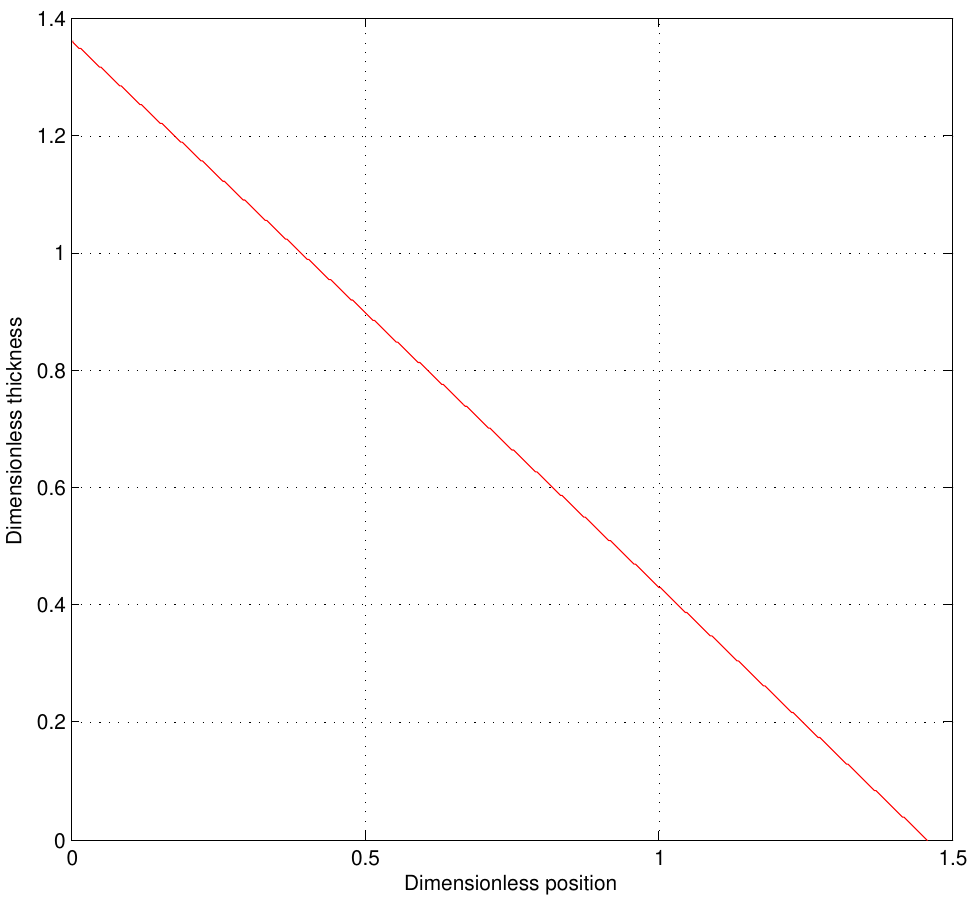}
  \caption{Dimensionless shelf thickness profile for $n = 3.8$. We promise we didn't just draw a triangle! The actual thickness of the shelf is found by scaling this using Equation \ref{Equation_41}.}
  \label{Similarity_Solution}
\end{figure}
%The figure needs to go on the previous page at all costs. If this is not possible, then the next line must definitely be after the profile.

Amazingly, the whole profile is very nearly triangular. However, the slope does steepen by about 3\% for $n = 3.8$. At higher $n$, this effect is reduced and both sides have length approaching $\sqrt{2}$. The reason is that $\psi \psi {{'}^{^{n}}}$ always goes from 1 to 0, and as $\psi \ne 0$ (except right at the front), we must have that 
\begin{eqnarray}
	\psi '\to 1 ~\forall \varepsilon \text{ as } n\to \infty 
\end{eqnarray}

We note briefly that for $n = 0$, the dimensionless profile is the unit square.

\begin{table} [ht]
\centering
\begin{tabular}{cccc}
\hline
$n$ & $\psi (0)$ & ${{\varepsilon }_{n}}$ & Fractional change in $u$\\
\hline
3.6	& 1.362	& 1.461	& 11.6\%\\
3.8	& 1.364	& 1.460	& 11.1\%\\
5.0	& 1.374	& 1.452	& 8.8\%\\
5.2	& 1.375	& 1.451	& 8.5\%\\
$\infty$	& 1.414	& 1.414	& 0\\
\hline
\end{tabular}
\caption{Results of computer simulations for those values of $n$ used in our experiments, and nearby values consistent with the error in $n$. Also included is the result for $n = \infty $.}
\label{Theoretical_profile_calculations}
\end{table}

%Values in the table need to be updated.

The expressions for the front position and the source thickness of the profile as functions of time are
\begin{eqnarray}
	{{x}_{n}}&=&\frac{{{t}^{\frac{n+1}{2n+1}}}{{d}^{\frac{1}{2n+1}}}{{Q}^{\frac{n}{2n+1}}}}{{{(n+2)}^{\frac{1}{2n+1}}}{{2}^{\frac{n-1}{2n+1}}}}{{\left( \frac{\rho g'}{{{\eta }_{o}}} \right)}^{\frac{n}{2n+1}}}{{\varepsilon }_{n}} \label{Front_position}\\
%\end{eqnarray}
%\begin{eqnarray}
{{H}_{0}}&=&\frac{{{t}^{\frac{n}{2n+1}}}{{(n+2)}^{\frac{1}{2n+1}}}{{Q}^{\frac{n+1}{2n+1}}}}{{{2}^{\frac{n+2}{2n+1}}}{{d}^{\frac{2n+2}{2n+1}}}}{{\left( \frac{{{\eta }_{o}}}{\rho g'} \right)}^{\frac{n}{2n+1}}}\psi (0)
	\label{E}
\end{eqnarray}

Notice that the gradient of the profile decreases with time (i.e. ${{H}_{0}}$ grows slower than ${{x}_{n}}$). This is to keep the entry flux the same despite a greater thickness (forcing a reduction in $u$ and thus $\left| H' \right|$).

As we have seen, thickening of the shelf is a hallmark of it being affected by viscous drag from sidewalls. For this to \emph{dominate}, we need to allow significant thickening of the shelf. However, at a length of $L$, it will only just have started to thicken, so sidewalls will only dominate when ${x}_{n} \gg L$. We expect convergence to be slow because it takes some time for the transient to die down (the decay is $\appropto {{x}_{n}}^{-1}$). This is because we assume that a section of shelf of length $L$ is unaffected by sidewalls, so it is outside our model (and creates something akin to a shift in position measurements). This region is essentially flat because the force balance was different when this region crossed the source (Figure \ref{Real_Profile}). The shelf behaves essentially as a solid body ($\frac{\partial u}{\partial x}$ is very small), so the result of this earlier time remains permanently imprinted upon the shelf. For our solution to work well, we need this region to be a very small part of the entire shelf. This way, the last vestiges of the times when sidewalls were unimportant will have faded into insignificance. 

Although we have estimated what length of shelf is required for sidewalls to dominate the system, this alone will not guarantee the similarity solution being accurate. This is because, even if the force balance was dominated by lateral friction from confining sidewalls, the amount of longitudinal stress \emph{inherent to our solution} could still be very large, making it internally inconsistent.

%Put in a du/dx term that varies with y in the usual manner, amounting to 0.1, 0.3 and 0.7 du/dy (at the edges). Consider effect on viscosity only. See if it makes much difference to q or to the ratio between du/dy at the edges and the mean velocity. When does it `feel like' a generalised Poiseuille flow? Use this to estimate the absolute minimum length of shelf (for a 15cm tank) that could plausibly appear to have converged. This should definitely be well below 30cm.

The difference in $u$ from the source to the front is approximately 10\% (for $n = 4$), so
\begin{eqnarray}
	\frac{\partial u}{\partial x} ~\approx ~ \frac{u}{10{{x}_{n}}}
\end{eqnarray}

Obviously, there will be a region close to the centreline of the channel where $\frac{\partial u}{\partial x} > \frac{\partial u}{\partial y}$, but as long as this region is small, our solution should be accurate. For this to occur, we compare $\frac{\partial u}{\partial x}$ along the centreline of the channel with $\frac{\partial u}{\partial y}$ at the sidewalls (i.e. we compare maximum values). For $n = 4$, this leads to the requirement that
\begin{eqnarray}
	\frac{u}{10{{x}_{n}}} ~\ll ~ \frac{4u}{d}
\end{eqnarray}
	
The conclusion, not altogether unexpected, is that the shelf needs to have a minimum aspect ratio. If we wish for $\frac{\partial u}{\partial y}$ close to the sidewalls to be at least $10$ times larger than $\frac{\partial u}{\partial x}$, then the shelf only needs to be half as long as the full width of the channel! Thus, the similarity solution is internally consistent for very short shelves. However, it does need to be much longer than $L$, and we believe that this is usually the stricter condition (it certainly was in our experiments).

We now touch briefly upon the effect of variations in thickness across the channel. In this case, a first integral of Equation \ref{BB} will no longer simply be directly proportional to $y$. Assuming the thickness is smaller near the sidewalls, then this will be a convex function. Therefore, $\frac{\partial u}{\partial y}$ will be greater than before, for the same average $H$ and $H'$. The effect of this can be determined by multiplying the formula for $q(x)$ by a factor greater than 1.

%The computer refuses to link to Equation BB, which you will just need to find manually from my labels.

However, the effect on the position of the front will be smaller than it might at first appear. Although the front must be further ahead than without the lateral thickness variation, this will also reduce the thickness and (combined with higher ${x}_{n}$), will reduce $\left| H' \right|$. Therefore, the fractional increase in ${x}_{n}$ (at the same value of $t$) is only $\frac {1}{2n+1}$ times as much as the fractional change in $q$. Thus, as long as the sidewalls are able to maintain the no-slip condition (i.e. as long as contact is not lost altogether), we expect the effect of lateral thickness variations on the front position to be small.

\newpage
\section{The Grounding Line}
\label{Grounding_line}

We now introduce a sloped bed at an angle of inclination of $\alpha $. The waterline is just above the top of this slope, with the weir just above the waterline. We now have both a sheet and a shelf, with the two linked at a grounding line. All parameters used previously still have the same meaning, except $d$. This is once again used for the full width of the shelf. A$\ {}_{G}$ subscript is used to denote parameter values at the grounding line.
 
\begin{figure}[ht]
  \centering
  \includegraphics [width = 16.5cm]{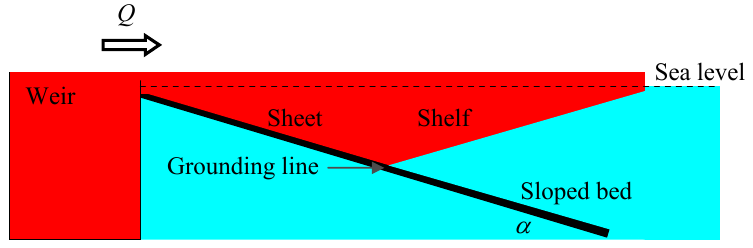}
  \caption{Side view of a channel flow with a grounding line.}
	\label{Grounding_Line_Setup}
\end{figure}

We model the grounded portion of the viscous layer (the sheet) as a viscous gravity current. We prove later that this is valid. We also believe it to be valid in most natural situations, but can not confirm this.

Assuming that $H \ll d$ (or that there are no sidewalls), we get that
\begin{eqnarray}
	\frac{\partial {{\sigma }_{xx}}}{\partial x} ~+~ \frac{\partial {{\sigma }_{xz}}}{\partial z} ~=~ 0
\end{eqnarray}

The assumption of a viscous gravity current is formally equivalent to approximating the $\frac{\partial {{\sigma }_{xx}}}{\partial x}$ term as a hydrostatic pressure gradient. Therefore, we get that
\begin{eqnarray}
	\frac{\partial }{\partial z}\left( \eta \frac{\partial u}{\partial z} \right) ~=~ \rho gh'
\end{eqnarray}

Solving the above equation subject to no slip at the base ($z = -b$) and a free upper surface ($\frac{\partial u}{\partial z}$ = 0 at $z = h$), we obtain the velocity profile for the sheet:
\begin{eqnarray}
	u(x, z) ~=~ {{\left( \frac{\rho gh'}{{{\eta }_{o}}} \right)}^{n}}\frac{{{2}^{1-n}}}{n+1}\left[{{H}^{n+1}}-{{(h-z)}^{n+1}} \right] \label{F}
\end{eqnarray}

The flux crossing a plane of constant $x$ is readily found to be:
\begin{eqnarray}
	q(x) ~=~ d{{\left( \frac{\rho gh'}{{{\eta }_{o}}} \right)}^{n}}\frac{{{2}^{1-n}}}{n+2}{{H}^{n+2}} \label{Q}
\end{eqnarray}

Note the similarity of the above equations with the corresponding ones for the shelf (with sidewalls). The confining surface runs parallel to the driving force, but in one case it is underneath and in another it is to either side. The shelf has no free surface like the sheet, but in the shelf the centreline of the channel acts as a free surface due to symmetry. Thus, we see that \emph{there is no fundamental difference between an ice sheet and an ice shelf confined by sidewalls}. Both have gravity balancing viscous drag and they also have similar boundary conditions, leading to a similar velocity profile.

\newpage
\clearpage
\subsection{No Sidewalls}

We solve first for the case where the shelf is not in contact with sidewalls. We assume that the flow does not spread laterally very much, or that it does so only over a very small region near the weir but not near the grounding line or in the shelf (so the width is constant in the regions we now discuss). We also assume that the grounding line has already reached its equilibrium position, so that conditions in the sheet close to this point are steady. In this case, we can set $q = Q$ near the grounding line.

The fundamental force balance at the grounding line is for the force exerted on the water-facing side of the fluid in the $x$-direction. This is because such a force can not be transmitted anywhere except into the shelf. Therefore, it must balance with the same force in the shelf (where it is created by hydrostatic pressure of the ocean). In other words, we require continuity of $\int_{-b}^{h}{{{\sigma }_{xx}}~dz}$ across the grounding line.

In the sheet, there is a contribution from hydrostatic pressure of $\frac{1}{2}{{\rho }}g{{H}^{2}}$. However, it is not balanced by the normal stress in the shelf $\left( \frac{1}{2}{{\rho }_{w}}g{{b}^{2}} \right)$. The difference is $\frac{1}{2}\rho g'{{H}^{2}}$. This must be accounted for by \emph{non-hydrostatic} forces in the sheet. Using the usual balance of vertical forces argument along with conservation of mass, we get that
\begin{eqnarray}
	I ~\equiv \int_{-b}^{h}{4\eta \frac{\partial u}{\partial x}\text{ }dz} ~=~ \frac{1}{2}\rho g'{{H}^{2}} ~\text{   at the grounding line}\label{G}
\end{eqnarray}
	
The pushing force $I$ is calculated directly from the velocity profile in Equation \ref{F}. We determine $I$ numerically given a particular value of $H$. The value of $h'$ is fixed by the requirement that conditions in the sheet near the grounding line be steady (i.e. $q = Q$).  Once $h'$ is found using Equation \ref{Q}, the computer next determines $\frac{\partial u}{\partial x}$ and $\frac{\partial u}{\partial z}$ as functions of $z$ at the grounding line, using also $H' = h' + \alpha$. The equilibrium thickness of the grounding line is then determined by varying $H$ so as to make Equation \ref{G} hold.

The viscosity is affected by both vertical and horizontal shear. Without horizontal shear, the integral will diverge for $n>2$, assuming a non-zero value of $\frac{\partial u}{\partial x}$ near the free upper surface. Thus, we use
\begin{eqnarray}
	\eta ~=~ {{\eta }_{o}}{{\left[ \sqrt{{{\left( \frac{\partial u}{\partial x} \right)}^{2}}+\frac{1}{4}{{\left( \frac{\partial u}{\partial z} \right)}^{2}}} \right]}^{\frac{1}{n}-1}}
\end{eqnarray}

The flow in the sheet is dominated by vertical shear, which vanishes at the free upper surface. For a shear-thinning fluid, the viscosity is thus greatest near this surface. Here, $u$ is also greatest. Thus, both $\eta$ and $\frac{\partial u}{\partial x}$ will be greatest here, so $I$ is only really affected by the value of $\frac{\partial u}{\partial x}$ near the upper surface.

%GRAPHS of sigma against z+b for n = 5, n = 1 and n = 0.3. Thickness half that which makes H' = 0.

The vertical velocity profile in the sheet has a thin boundary layer (for large $n$), so in order to have flux conservation we must approximately have that $u=\frac{Q}{Hd}$ outside this region. Thus, we expect that $I$ will change sign when $H'$ changes sign (i.e. when $h'=-\alpha$). At the corresponding value of $H$, $I$ should be very small.

Computer simulations indicate that, for $H$ fairly close to the `right' value but not exactly equal to it, $I$ is very sensitive to $H$. Thus, the value of $H$ which makes Equation \ref{G} hold and the value of $H$ which makes the integral 0 are often quite close. This is equivalent to saying that the pushing force can easily be made quite large compared with the hydrostatic pressure discontinuity. Thus, solving $I = 0$ is approximately the right thing to do. This suggests that the grounding line thickness may be approximated by assuming that $h' = - \alpha$ there, so that
\begin{eqnarray}
	{{H}_{G}} ~\approx ~ {{\left[ \frac{Q(n+2)}{d} \right]}^{\frac{1}{n+2}}}{{\left( \frac{{{\eta }_{o}}}{\rho g\alpha } \right)}^{\frac{n}{n+2}}}{{2}^{\frac{n-1}{n+2}}}\label{H}
\end{eqnarray}

Notice that $g'$ is irrelevant if this approximation is accurate. 

\newpage
\begin{figure}[ht]
	\centering
		\includegraphics [width = 14cm] {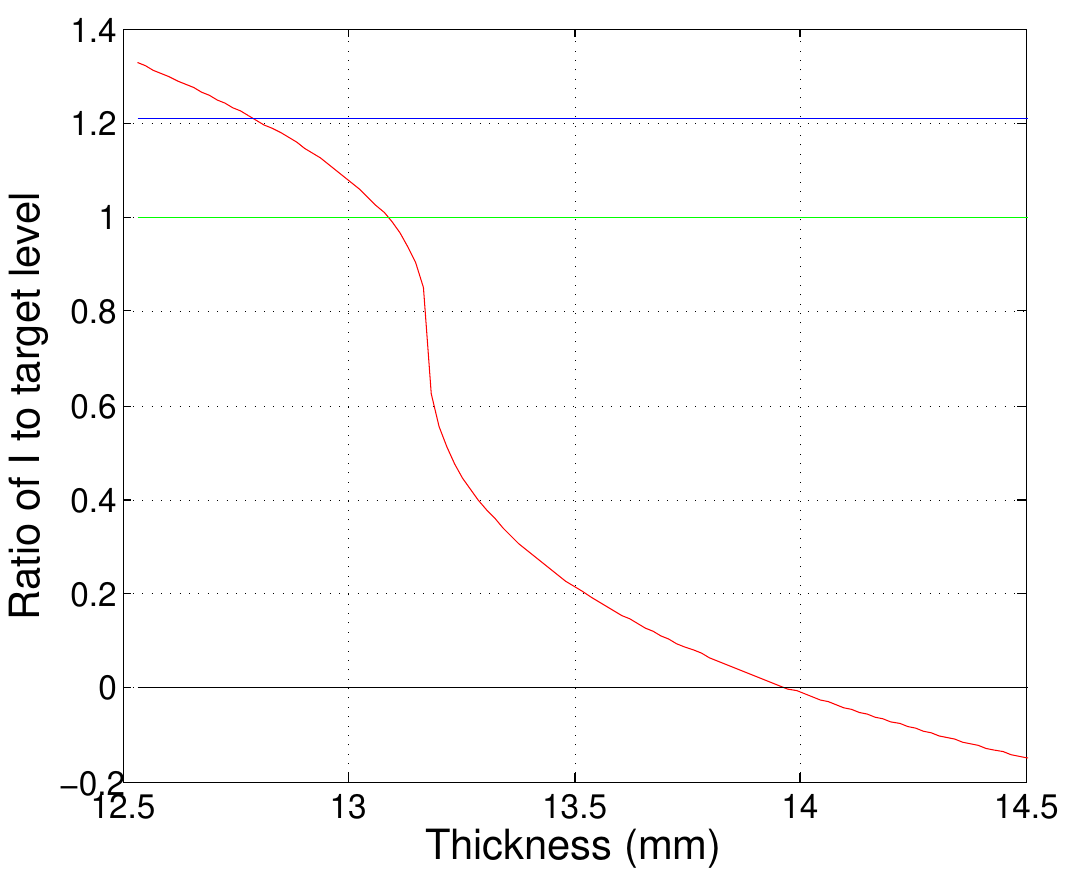}
	\caption{For parameter values matching those of one of our experiments, $I$ divided by the hydrostatic pressure discontinuity is shown as the red curve. The green line is at 1. However, the intersection of the blue line with the curve is a good approximation. The location of this point is given in Equation \ref{H}.}
	\label{Grounding_Line_Model}
\end{figure}

Equation \ref{H} requires $H' = 0$, but it does not actually result in $I = 0$. Because conditions remain the same if we move parallel to the sloped bed, moving along $x$ \emph{at fixed $z$} will lead to a geometric effect whereby
\begin{eqnarray}
	\frac{\partial u}{\partial x} ~=~ \alpha \frac{\partial u}{\partial z}
\end{eqnarray}

Thus, the value of $I$ divided by the hydrostatic pressure discontinuity is not exactly $0$. It is
\begin{eqnarray}
	{{2}^{1+\frac{1}{n}}}{{\alpha }^{2}}{{\left( {{\alpha }^{2}}+\frac{1}{4} \right)}^{\frac{1}{2}\left( \frac{1}{n}-1 \right)}}\frac{g}{g'} \label{Equation_59}
\end{eqnarray}

If this is close to 1, then Equation \ref{H} will be a good approximation to the result of a full computer simulation (and the ice will be running nearly parallel to the sloped bed). A ratio above 1 means that Equation \ref{H} underestimates ${H}_{G}$ (as in \text{Figure} \ref{Grounding_Line_Model}). For ice in sea water, we note that the equation holds exactly when $\alpha = 9^\circ$. 
%The high accuracy achieved by us meant that, for our experimental parameters, the approximation gave a small fractional error but one which exceeded observational errors. 

We warn the reader to check this ratio and the results of the computer simulation, to see what sign the error resulting from using Equation \ref{H} is and whether its magnitude is acceptable. For now, the system is sufficiently simple that the full simulation only takes a few minutes. In more complicated systems, having a simple equation for the grounding line thickness may prove to be valuable, even if it is inexact.

Note that the green line will appear closer to the black line at very low $g'$ (and further for higher $g'$), whereas the blue line will appear not to move. $g'$ does matter. As expected, a less buoyant fluid will have a thicker grounding line. Interestingly, though, reductions in $g'$ can not raise the grounding line thickness above a certain value (although increases in $g'$ can lower $H_G$ without limit).

\newpage
\clearpage
\subsection{With Sidewalls}
\label{Grounding_line_sidewalls}

%This page is for guidance only and certainly not to be published like this. It is just my idea about what must be going on, but we already solved the late-time case - it's exactly the same as before! Experiments confirm this.

The essential difference in the presence of sidewalls (assuming they dominate the system) is that, once a shelf with a particular grounding line thickness is formed, there \emph{is} a tendency for this thickness to change. Sidewall friction causes fluid to essentially `pile up' behind the front to some extent, not just to flow completely freely as it does in the case of no sidewalls.

This `piling up' means that there is \emph{no} stable grounding line thickness. Therefore, the shelf thickens for ever. However, there is still a dynamic balance at the grounding line. This is because if all the flux entered the shelf, then it would want the grounding line thickness to increase at a certain rate. However, the sheet retains no flux, so it can't grow. Thus, the grounding line can not advance. 

The impossibility of the situation reveals what must really happen: part of the flux is retained by the sheet, allowing the grounding line to advance; while part goes into the shelf, presumably an amount equal to that which causes ${{H}}_{0}$ to increase by precisely the rate at which the flux retained by the sheet allows. This means that there is a balance between dynamic conditions in the shelf (how much it wants to thicken, given the flux entering it) and kinematic conditions in the sheet (how much it must expand, given that it retains the flux not entering the shelf).

Eventually, the flux entering the shelf approaches Q. This is because the flux retained by the sheet is approximately equal to $H{}_{G}\text{ }\overset{.}{\mathop{{{x}_{G}}}}\,$, where a time derivative is indicated. Of course, for a fixed angle sloped bed we have that $\overset{.}{\mathop{{{x}_{G}}}}\,\propto \overset{.}{\mathop{{{H}_{G}}}}\,$. Considering that ${{H}_{G}}\propto {{t}^{\frac{n}{2n+1}}}$, we see that ${{H}_{G}}\text{}\overset{.}{\mathop{{{H}_{G}}}}\,\propto {{t}^{-\frac{1}{2n+1}}}$. Thus, the flux retained by the sheet inevitably goes down to 0, but fairly slowly. This means that, even with a grounding line, the shelf will eventually converge to the similarity solution we found earlier (whether we consider the length of the shelf only, or the position of the front). The slow convergence may mean that in a real laterally confined ice shelf, it needs to be fairly long in order for all the flux to enter the shelf.

Ultimately, if one is interested in what happens before convergence has occurred, a computer simulation will be required. This will need to solve our non-linear diffusion equation for the shelf and a similar version for the sheet. The boundary condition must be that flux not entering the shelf is retained by the sheet (and may go into causing grounding line advance). Similar models have already been devised for Newtonian fluids \cite{Pegler_2013, Kowal_2016}.

We assume that the grounding line rapidly reaches a thickness such that $\left|h'\right|\ll \alpha$, so that $H'\approx \alpha$. This lets us approximate that
\begin{eqnarray}
	\frac{\partial u}{\partial x} ~\approx ~ -\frac{Q}{{{H}^{2}}d}\alpha 
\end{eqnarray}

As there will be something like an extra power of $H$ in the total pushing force exerted by the sheet (to account for the vertical integration), we see that this scales with time inversely to $H$.

In the shelf, we have that 
\begin{eqnarray}
	\frac{\partial u}{\partial x} ~\approx ~ \frac{Q}{Hd\text{ }{{x}_{n}}}
\end{eqnarray}

Hydrostatic pressure of the ocean is of course completely dissipated by sidewall friction. The pushing force in the shelf will also need to have an extra power of $H$, so this scales with time inversely to ${{x}_{n}}$. We expect this to grow faster than $H$, on the basis of our similarity solution for the shelf (which the system converges to, eventually). Thus, in the end, the pushing force from the sheet will always exceed that from the shelf.

The force balance at the grounding line still needs to hold. Now, ${\sigma}_{xx} \equiv - P + 2 \eta \frac{\partial u}{\partial x}$, so $P$ needs to be greater in the shelf than in the sheet. We believe this to mean that there is a sharp increase in $H$ immediately after the grounding line, with this sudden change in $H$ accounting for the discrepancy in the vertically integrated pushing force that we have just found. However, the effect becomes negligible in the asymptotic limit. We never noticed such an effect in any of our experiments, suggesting that it may be irrelevant.

\newpage
\section{Experiments}
\label{Experiments}
\subsection{The setup}

\begin{figure}[ht]
  \includegraphics [width = 15.5cm]{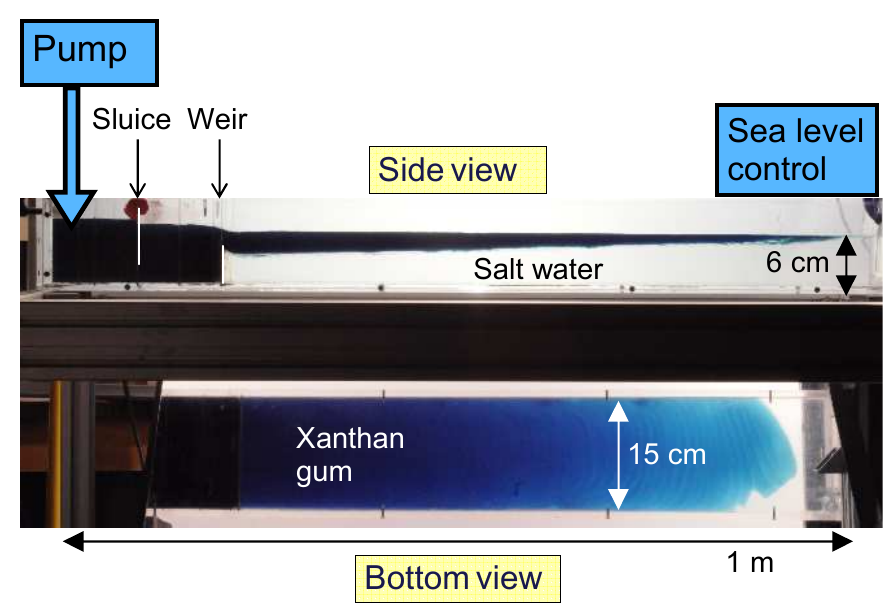}
  \caption{The experimental apparatus used (distances except the 15 cm channel width are approximate). The bottom view is acquired using a mirror at $45^\circ$ to the horizontal. A sloped bed was sometimes installed immediately after the weir (as shown in \text{Figure} \ref{Grounding_Line_Setup}).} 
	\label{Experimental_Apparatus}
\end{figure}

The basic apparatus is shown in \text{Figure} \ref{Experimental_Apparatus}. A peristaltic pump was used to maintain a constant flux into the region behind the sluice. The viscous fluid used was an aqueous suspension of xanthan gum at concentrations of 0.5\% and 1\% by mass. Xanthan is a shear-thinning organic polymer. Salt was added to the ocean to increase its density.

The viscous fluid was overflowing the weir and dropping, creating a rebound effect. We minimised this by leaving a very small gap between the ocean level and the top of the weir. We sometimes wished to include a sloped bed in order to study grounding lines. We found that leaving even a tiny part of the slope exposed to air had a dramatic and adverse impact upon our experiments, as occurred in some previous investigations \cite{Robison_2010}. We therefore decided to have it entirely submerged. To guarantee the formation of a sheet and also to reduce the rebound effect just mentioned, the sloped bed was usually placed 2 mm below the top of the weir. The sea level was kept halfway between the top of the slope and the top of the weir, as indicated in \text{Figure} \ref{Grounding_Line_Setup}. Maintaining this configuration required accurate control of the sea level.

%Include photograph of ripples? I've got a close-up of fluid going past the weir, this one told us that fluid overflows at 30 degrees. Obviously, you can get this too by zooming in to most photographs. Anyway, we dealt with the issue, though the 30 degree angle might be interesting.

We achieved this by means of a laser reflecting off the ocean surface onto a fixed screen. Water was siphoned out at a variable rate, with manual adjustments to this rate whenever necessary to keep the laser spot at the same location on the screen. We found that the rate of seawater extraction could be altered by 0.2 g/s, so we could easily control it accurately enough for our purposes. It is likely that the sea level was controlled to within 1 mm for most experiments, and 0.5 mm for some of them where there were less ripples on the water (usually due to a lower flux). Therefore, systematic trends in the sea level were small during all of our experiments.

The most probable cause of errors is simply the strong sensitivity of the experiments to initial conditions. Thus, a slight asymmetry (e.g. due to the tank being slightly tilted to one side) can cause loss of contact with the sidewalls at early times, leading to the sort of pattern seen in \text{Figure} \ref{Experimental_Apparatus}. The finite extent of the experiments also created a finite error on any experimentally determined power law dependence of the parameters on time. This was mostly due to difficulties in determining precisely when the experiment started. Because xanthan would not usually overflow the whole weir at the same time, the flux entering the channel would rise from 0 to $Q$. This would cause the front to accelerate. To overcome this problem, we usually waited for the front to stop accelerating and then did a regression of front position against time. We took the time at which this regression line passed through 0 cm to be the time origin for the whole experiment. 

Our theory for no sidewall experiments indicates that acceleration not due to rising flux, if present at some time, must also necessarily be present at all later times during the experiment. As the front only accelerated at early times during these experiments, we concluded that this was in fact due to changing flux, and so this must also be the case for similar experiments with additional viscous drag from sidewalls. Thus, all instances of the front accelerating are ascribed to a known artefact of our experimental setup.

Although we consider the procedure perfectly reasonable, it does lead to an error of at least 1 second on the time origin used for the whole experiment, and sometimes as much as 10. Also, the shelf was not usually thickest at the weir itself, but a few cm beyond it. Upstream of here, bending moments were likely having a significant impact upon the flow. As such forces are outside our model, they lead to the model only becoming valid for regions downstream of the point of maximum thickness. This leads to the conclusion that all position measurements should be relative to the point where bending moments become insignificant. Of course, the location of this point has an error of about 1 cm (though for very thick shelves, it may be much more). For consistency, though, we would also need to subtract the time required to fill up the section of the profile behind this point. For simplicity, we did neither, believing the effects to be roughly comparable and fairly small in any case. Experiments where this was not so are excluded from our analysis (though we show the data anyway).

Because our procedure essentially assumed no flux entering the channel at all until such time as all $Q$ was entering the channel, we underestimate the amount of fluid in the channel. This means the experiment effectively got underway earlier than we are assuming, causing us to overestimate the intercepts and underestimate the gradient on the log-log graphs we used. One possible solution is to accurately determine the total amount of fluid which crossed the weir. This could be done if we knew that the sea level had been maintained very accurately and measuring how much water had to be extracted to achieve this. As soon as the experiment finished, the peristaltic pump could be reversed, to prevent further flow of xanthan into the ocean. Then, the position of the laser spot for sea level control could be compared with its initial position to see how much change there had been. We estimate that the total volume of fluid pumped into the ocean could be determined to within 40 cm$^{3}$, corresponding to an error of only a few seconds on the effective start time. However, not realising the importance of it, we did not perform this procedure.%These are also excluded from our analysis, but again the data is shown.

Another source of uncertainty was the wavelike oscillations that are evident in the bottom view above. These are due to hydrostatic rebound of xanthan dropping into the ocean from a finite height. We tried various techniques to reduce the effect of such waves, and were successful in nearly eliminating them. Therefore, only two experiments were significantly affected by this phenomenon.

There are a few more sources of error worth mentioning. Firstly, the concentration of xanthan may have been slightly below 0.5\% because of losses in transferring the powder from the container in which it was weighed into the water. We assume that $\ssim 5\%$ of the xanthan may have been lost in this way, meaning the concentration may have been systematically lower (only 0.47\%, say). This will reduce $\eta_o$ by about 10-15\%. Also, the shear rates in our experiments may have been sufficiently low that the power law used to model the viscosity (Equation \ref{Viscosity_law}) breaks down. Ultimately, the viscosity is not infinite at very low shear rates, so the fluid must be less viscous than we are assuming.

Fluxes were measured by weighing the container from which xanthan was pumped into the tank. Although a slightly different amount may have been overflowing the weir and entering the shelf (especially near the start of our experiments), this is only true for a very short time. The measurements of mass flow rates were very accurate ($\ssim 0.1\%$).

The density of the ocean was measured using a hydrometer, attaining a similar level of accuracy. For xanthan, we put it into saltwater of known density and checked if it floated or sank. When 50\% of the samples we put into the water sank, we knew we had the right density. We also checked this using a hydrometer. Both gave consistent results, with an accuracy of $\ssim 0.1\%$ on the density. This corresponds to an error of under 2\% on $g'$. Our density measurements are listed below. We simply extrapolated the density of xanthan at 1\% concentration to be 996 kg/$m^3$.

\begin{table} [ht]
  \centering
  \begin{tabular}{cc}
	\hline
    Substance &	Density (kg/$m^3$)\\
	\hline
	Water	& 994\\
	Xanthan at 0.5\%	& 995\\ \hline
	\end{tabular}
\end{table}

The position of the front as a function of time was determined by a \textsc{matlab}$^\text{\textregistered}$ boundary tracing algorithm. The positions are listed relative to the weir if there was no sloped bed, relative to the point of maximum thickness if this was more than 3 cm from the weir and relative to the location of the grounding line if there was a sloped bed. However, experiments in which the position of maximum thickness was more than 3 cm from the weir were severely affected by bending moments, making the results of these experiments unusable.

We then saw if the slope of a graph of ${x}_{n}$ against $t$ (on logarithmic axes) converged. This is done by requiring the residuals to a linear regression (usually below 0.5\%, and sometimes just a tenth of this) to not have a characteristic inverted parabola shape, but to appear essentially random. We list the portion of the tank over which this occurred, and the relevant times. Also listed is the product-moment correlation coefficient, to give an idea of how closely the data fit to a straight line. If an experiment did not converge, then the gradient would still be decreasing by the end of the experiment (because the gradient needs to go down from 1 to $\ssim 0.6$). In this case, we did a regression on the last $\ssim 30$ seconds of data, to give a bound on what the gradient might eventually converge to, as well as where the intercept could then lie. This equates to an upper bound on the final gradient and a lower bound on the intercept.

%Include a graph where the gradient is 0.6-0.01/t and 0.6 + 0.01sin t (to model scatter). They look the same to the eye, but then get a line of best fit done and plot the residuals.

Usually, we also excluded data taken in the last $\ssim 5$ cm of the tank, to allow for the effect of the sea level control mechanism on the shelf. If there was no discernible effect, we used the additional data in our regression. Ocean currents can affect the xanthan because water has a finite viscosity. The effect is almost always to cause a sudden increase in the gradient (on logarithmic axes). However, for very high fluxes, we believe that the change in water pressure favours thickening of the shelf and thus slows it down even further. 

If the reader is interested, we strongly recommend manually analysing the (few) photographs from the very end of an experiment (at 8 g/s and at 17 g/s, to see both regimes). Another interesting thing to try is to exclude the possibility that the effect near the end is part of a long-period wavelike oscillation (we damped these, but they might still be present). This is relatively straightforward $-$ the experiment simply needs to be repeated with the weir moved forwards 10 cm or so. That way, the end of the tank would correspond to a different phase of the (hypothetical) wave. We also note that a much more viscous ocean (e.g. using sugar rather than salt to reach the target ocean density) would enhance the effect.  However, it was not our intention to understand the influence of ocean currents on ice shelves, so we do not discuss this further.

Our experiments are given between one and three letters and a number to help the reader identify and refer to them. The letters indicate respectively the presence of laterally confining sidewalls, the presence of a sloped bed and the concentration of xanthan used for the experiment. 

\begin{table} [ht]
  \centering
  \begin{tabular}{cc}
	\hline
    Letter &	Meaning\\
\hline
W	& Sidewalls\\
B	& Bed\\
H	& 1\% concentration used\\
L	& 0.5\% concentration used\\ \hline
		\end{tabular}
\end{table}

%Can we please have lines around everything, like if the table was in Word?

A typical experiment will be identified by e.g. L1 (indicating no sidewalls, no sloped bed and a concentration of 0.5\%). The number is self-explanatory. If these do not start at 1 or miss a number, this is because an experiment was excluded from this paper. In this case, a good reason will be given.

Finally, we note that only an error in the concentration of xanthan used (and thus in $\eta_o$) still remains as a systematic effect in the experiments mimicking ice tongues. Other errors for these experiments are purely random, the biggest of which is in measuring the width of the shelf. The thin parts near the edges had to be excluded from our measurement of $d$, for reasons that will become apparent. Such a procedure inevitably creates some error and is partly subjective.

\newpage
\clearpage
\subsection{1\% aqueous xanthan solution}
\label{One_percent_xanthan}

Experiments with sidewalls were all performed in the same tank with $d = 0.075$ m (as it was manufactured, the error is negligible). We used ${{\rho }_{w}} = 1100$ kg/$m^3$ for all WH experiments. Experiments WH1-3 are not included because we were still perfecting our techniques and because the weir had some rust. This severely hampered our experiments because xanthan overflowing the weir tended to stick to the rust rather than flow forwards into the ocean. When the xanthan finally left the weir, it had gone a long way down so there was a huge blob at the front of the shelf. We warn readers attempting to repeat our experiments that they are of an extremely sensitive nature, especially those without sidewalls.

We do not include two experiments conducted at an extremely low flux. This is because the shelf was so thin that it lost contact with the sidewalls at a very large number of locations. There was also insufficient sidewall contact to make one experiment converge, although it suggested that the power of $t$ is $< 0.56$. It also suggested a higher intercept than other experiments, although this is almost certainly due to the loss of contact with sidewalls (which reduces the drag on the shelf).

\begin{table}[ht]
	\centering
		\begin{tabular}{ccccccccc}
\hline
Expt. & Flux & Convergent & Error	& ${R}^{2}$	& Time (s) & Distance & Intercept & Error\\
(WH..) & (g/s) & power of t	& & & & (cm) & & \\	
\hline
10	& 6.23	& 0.554	& 0.01	& 0.9997	& 216-239	& 66-70	& $-3.38$	& 0.06\\
9	& 12.41	& 0.540	& 0.01	& 0.9997	& 101-146	& 56-68	& $-3.08$	& 0.06\\
7	& 15.18	& $< 0.61$	& & & &	&	$> -3.30$	& \\
6	& 7.87	& 0.541	& 0.01	& 0.9997	& 163-189	& 60-66	& $- 3.21$	& 0.06\\
4	& 3.87	& $< 0.61$ & & & & &	$> -3.78$ & \\ \hline			
		\end{tabular}
	\caption{The results obtained for our experiments with $1\%$ aqueous xanthan solution.}
	\end{table}

The experiments which did converge were all consistent with each other. Our best estimate for the mean value of the convergent power of $t$ is
\[0.545\pm0.006\]

 If ${x}_{n} \propto {t}^{\frac{n+1}{2n+1}}$, as predicted by our theory, then we require a value for $n$ of
\[n~=~{5.1}_{-0.7}^{+0.8}\]

This is entirely consistent with the value of $\ssim 5$ suggested by independent measurements of $n$ for this fluid \cite{Taylor_2003}.

%Give references and check the numbers.

% As already shown, only a X\% difference between mean and edge thicknesses can lead to a $\sim Y\%$ discrepancy between the predicted and measured values of $u$ (and thus of ${{x}_{n}}$).

Next, we check if the intercepts are also consistent with our theory. However, we should not expect them to be. This is because it was obvious that there are significant thickness variations across the channel. Considering how easy it was for the shelf to lose contact with the sidewalls altogether, we suppose that there must have been a significant variation in thickness across the channel. Presumably, the wider the channel or the more viscous the fluid, the more difficult it is to transport mass towards the sidewalls. This is essential to making these regions thicken with time (along with the rest of the shelf). We thus predict that a narrower channel should give better agreement with our theory, as should a less viscous fluid.

We now proceed to rescale the intercepts (on a log-log graph) based on changes in $Q$, remembering that $x_n \propto {Q}^{\frac{n+1}{2n+1}}$. Once the rescaling is done, the intercepts should (theoretically) all be equal.

\begin{table}[ht]
	\centering
		\begin{tabular}{cccc}		
\hline
Expt. 	& Flux 	& Rescaled 	& Error\\
(WH) & (g/s) & intercept & \\
\hline
9	& 12.41	& $-4.22$	& 0.06 \\
10	& 6.23	& $-4.21$	& 0.06 \\
7	& 15.18	& $> -4.5$	& \\
6	& 7.87	& $-4.14$	& 0.06 \\
4	& 3.87	& $> -4.4$	& \\ \hline
		\end{tabular}
\caption{The intercepts obtained for the experiments with $1\%$ xanthan solution, rescaled according to our theory and the alterations in flux. The theoretical value is -4.99, assuming $n = 5$ and ${{\eta }_{o}} = 10$ Pa ${s}^{\frac{1}{n}}$.}
\end{table}
%Please check the theoretical calculation I did - this appears to give -4.99 (i.e. Ln(x_n) = -4.99 + 0.56Ln(t) in SI units), if flux rescaled to 1g/s (or Q = 1.004cm^3/s), g' = 1m/s^2 and other parameters as you expect. However, E_N needs to be checked.

As can be seen from Table 3, the values are roughly consistent, although the theoretical value is about -5. Our best estimate for the rescaled value of the intercept is
\[-4.19\pm 0.04\]
%State theoretical value here. I think it's -4.99.

The discrepancy with our theory could be due to lateral thickness variations and partial loss of sidewall contact throughout the shelf. The scaling of $x_n$ with $Q$ appears to be as expected, but the changing relative importance of lateral thickness variations leads to this not being completely correct either. The net effect is that experiments at a lower flux go slightly faster than we would predict from scaling data for a higher flux experiment. Presumably, this is due to experiments with lower $Q$ leading to lower $H$, making sidewall contact less likely.

Another interesting thing to note is that the impact of lateral thickness variations was very similar for all experiments. This suggests that the fractional variation in thickness across the channel was much the same, so the lateral thickness profile might be self-similar inside and between experiments. Otherwise, the data would not remain parallel to our theoretical solution. Future work may elucidate this further.

The intercepts are given when the data (in SI units) is plotted on a log-log graph (with base 10). The rescaled intercepts are what would be obtained if the scaling predicted by our theory is correct and the flux was reduced to 1 g/s (so $Q = 1.004~{cm}^{3}$/s).

%Confirm that the density of 1% xanthan is 996kg/m^3, using the neutral buoyancy in an inviscid fluid method or another one of your choice, but DON'T PRETEND YOU KNOW (though it's 995 for 0.5% and water is 994).

\begin{figure}[ht]
	\centering
		\includegraphics [width = 16.5cm] {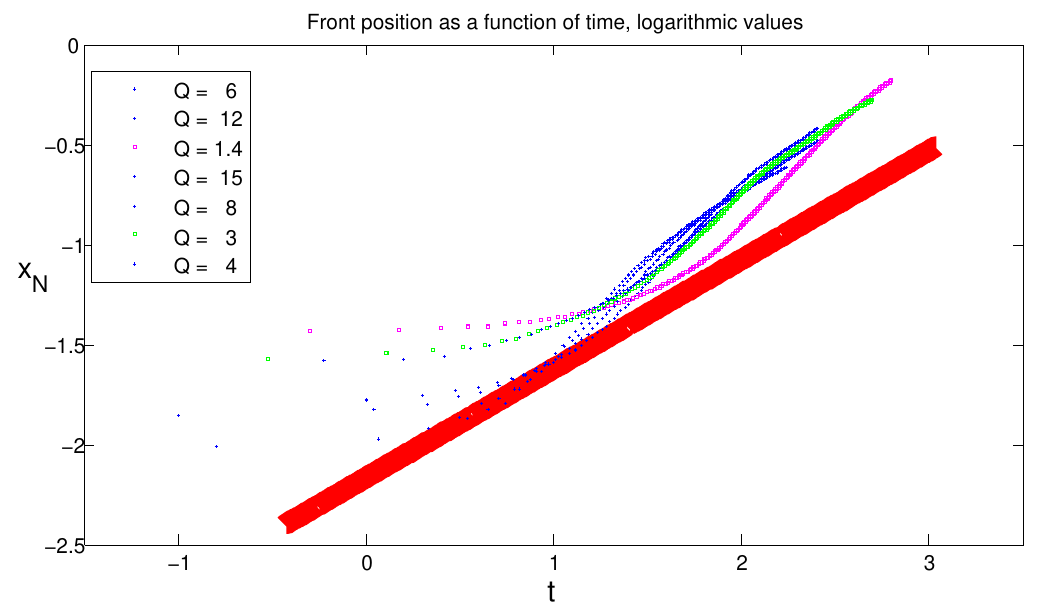}
	\caption{Rescaled front positions as a function of time on logarithmic axes for all reliable WH experiments (the most reliable ones are shown in blue). Experiments at very low fluxes had too much lateral thickness variation to be considered reliable, and they showed the strongest disagreement with our asymptotic theory (Equation \ref{Front_position}). All experiments lie well above the theoretical line, shown in red. Surprisingly, though, they differ from the theoretical curve by a very similar amount, suggesting the scaling relations with $Q$ and $t$ still work.}
	\label{One_Percent_Data}
\end{figure}

\newpage
\clearpage
\subsection{0.5\% aqueous xanthan solution}
\label{Half_percent_xanthan}

We attempted to minimise the effects of lateral thickness variations upon the shelf. This can be done by reducing the width of the shelf, but we attempted instead to reduce the viscosity of the fluid by reducing the concentration of xanthan to 0.5\%. This reduces the viscosity by a factor of $\ssim 3$. It also reduces $n$, which is highly desirable as ice has $n \approx 3.2$ at the temperatures of geophysical interest \cite{Glen_1955}. However, we did not quite reduce the concentration enough to reach this, because if we did then the fluid would not be very viscous and the flow might fail to be at low Reynolds number.

\begin{table}[ht]
	\centering
		\begin{tabular}{cccccccccc}
\hline
Experiment & ${\rho}_{w}$ & Flux & Convergent & Error & ${R}^{2}$ & Time (s) & Distance & Intercept & Error\\
(WL..) & & & power of t & & & & (cm) & & \\		
\hline
1	& 1100 & 6.62	& $<0.61$ & & & & & $>-3.67 $	&\\
2	& 1100 & 12.05 & 0.56	& 0.03 & 0.9994 & 47-163 & 37-75 & $-3.14$ & 0.1\\
3	& 1100 & 16.73 & 0.53	& 0.03 & 0.9997 & 66-123 & 51-71 & $-2.9$ & 0.2\\
5	& 1100 & 3.96 & 0.58 & 0.02	& 0.9985 & 309-350 & 64-69 & $-3.77$ & 0.06\\
B6 & 1053 & 11.79	& 0.57 & 0.02	& 0.9986 & 105-233 & 42-66 & $-3.52$ & 0.06\\
7	& 1053 & 12.29 & 0.61	& 0.03 & 0.9993	& 68-213 & 41-82 & $-3.48$ & 0.06\\
B10	& 1029 & 3.9 & 0.56	& 0.015 & 0.9842	& 420-684	& 44-57	& $-4.24$	& 0.06\\
B11	& 1100 & 3.79 & $<0.61$	&	& & & &	$>-4.24$	&\\
B12	& 1100 & 15.5	& $<0.80$	&	& &	&	&	$>-4.26$	&\\
B13	& 1029 & 8.12	& 0.56 & 0.015 & 0.9946 &	175-387	& 39-58	& $-3.88$	& 0.06\\
B14	& 1053 & 15.9	& $<0.67$ & & & & &	$>-3.66$	&\\  \hline
\end{tabular}
\caption{The results obtained for our experiments with $0.5\%$ aqueous xanthan solution.}
\end{table}
%Please redo the error calculations. I think the Excel file has more updated values. Basically, `good' experiments should have 0.06 for the intercept. Error in gradient about right.

We have not included two experiments which had extremely thick shelves, due to bending moments near the weir playing a role over a large section of the tank. Most likely, the correct thing to do is to subtract $\ssim 10$ cm from the front positions to account for this region (as our theory only becomes valid beyond it). This is also suggested by the fact that, unlike all other experiments, the gradient on a log-log graph (of position against time) was actually increasing (rather than decreasing from 1 at early times towards $\ssim 0.5$), strongly suggesting a zero error. However, because we could not precisely determine what correction to use, we do not include such experiments. 

Also not included is the first experiment we conducted that had a sloped bed. This was partially exposed to air, which led to an unusual start to the experiment and loss of contact of the shelf with the sidewalls over a 5 cm region near the front. Subsequent experiments had a much shallower ($\ssim 10^\circ$ instead of $26 ^\circ$) sloped bed installed, as well as this being entirely submerged. Contact with the sidewalls was much improved as a result.

Unsurprisingly, the additional length of tank used up by the sheet meant that convergence to self-similar propagation was harder to obtain (although it greatly improved the quality of the experiment). However, we still managed it on three occasions. As the sloped bed did not appear to have a significant effect on the shelf, we treat \emph{all} WL experiments in the same way.

Errors were raised slightly by the lower viscosity of the fluid, which made it more prone to oscillations due to hydrostatic rebound. However, experiments with a sloped bed or with a flux below 7 g/s were only slightly affected. This meant that only experiments WL2 and WL3 are noticeably affected. In what follows, we do not use either because we could not average over enough oscillations.

Our best estimate for the asymptotic behaviour of the shelf is that the front propagates as a power law in $t$ with exponent
\[0.565\pm 0.008\]
	
This implies a value for $n$ of
\[n ~=~ {3.3}_{-0.4}^{+0.6}\]

%Check the values!
Getting independent values for $n$ proved difficult. In the end, we found values at lower concentrations of xanthan than we used, and extrapolated them to 0.5\%. The value of $n$ at 0.4\% was 3.33, and at 1\% it is close to 5. Also, at 0.2\% it is 2.83. Thus, we expect that at 0.5\% $n$ should be about 3.8, making it consistent with our observations. The reference we used was \cite{Taylor_2003}: \newline

\url{http://projekt.sik.se/nrs/conference/Old\%20conferences/conf2003/Course2003/course\_Taylor\%20.pdf}
\newline

We obtained a value for ${{\eta }_{o}}$ in a similar way. At 0.2\%, it is 0.57 Pa ${s}^{\frac{1}{n}}$. All values of ${{\eta }_{o}}$ are given in these units. At 0.4\%, it is 2.24 and at 1\% it is about 10. Thus, we expect a value of about 3.5 at 0.5\%. This allows us to check whether the intercepts are also consistent with our theory.

The values of ${{\eta }_{o}}$ and $n$ could be verified by e.g. pumping the fluid into a dry Hele-Shaw cell, forming a lateral-shear dominated viscous gravity current. This is already a well-understood situation, so advantage could be taken of this fact. Other more advanced techniques are also possible.

%Actually do this! Then, update the theoretical value of -4.45.

\begin{table}[ht]
	\centering
		\begin{tabular}{cccccc}		
\hline
Experiment & Ocean & Flux & Rescaled & Error\\
(WL..) & density & (g/s) & intercept\\
\hline
1 & 1100 & 6.62 & $ > -4.45$ &\\
2 & 1100 & 12.05 & $ -4.18$ & 0.1\\
3 & 1100 & 16.73 & $ -4.07$ & 0.2\\
5 & 1100 & 3.96 & $ -4.32$ & 0.06\\
B6 & 1053	& 11.79 & $ -4.31$ & 0.06\\
7 & 1053 & 12.29 & $ -4.29$ &0.06\\
B10 & 1029 & 3.9 & $ -4.33$ &0.06\\
B11 & 1100 & 3.79 & $ > -4.8$	&\\
B12 & 1100 & 15.5 & $ > -5.4$	&\\
B13 & 1029 & 8.12 & $ -4.29$ & 0.06\\
B14 & 1053 & 15.9 & $ > -4.6$	&\\ \hline
\end{tabular}
\caption{Intercepts on a log-log graph obtained with $0.5\%$ xanthan experiments. These are rescaled so as to be what one would get for an experiment at 1 $g/s$ and with $g' = 1$ m/${s}^{2}$ if using SI units for other quantities and base ${e}$. The theoretical value is $\sim -4.42 \pm 0.05$, the error arising due to uncertainties in interpolating the viscosity of xanthan to the concentration used.}
\end{table}
%Please recalculate theoretical value with better data on the viscosity parameters.

The mean value for the intercept that we obtain is:
\[-4.31\pm 0.03\]

This is slightly greater than the theoretical value of $-4.42$. We have already mentioned one possible cause of the discrepancy $-$ errors in the start time. However, although it is a systematic effect, its magnitude will likely be below the error budget quoted above. An error in the concentration of xanthan could also be to blame (a 15\% reduction in $\eta_o$ leads to a 7\% increase in ${x}_{n}$), as could the low shear rates in the experiments. Lateral thickness variations could also account for a further 2\% discrepancy. Temperature could plausibly be a factor as well $-$ the lighting we used was very inefficient and could definitely have heated a dark fluid. Combining all of these effects, the relatively small discrepancy between theory and observations could conceivably be explained.

To test these ideas, the viscosity of the fluid we used, prepared in the same way as for the above experiments; should be measured directly. One possibility is to use a very narrow tank (a Hele-Shaw cell) and have no ocean, using the viscous gravity current theory to determine the viscosity parameters. Also possible is to use the technique outlined to get a better estimate for the start time. Repeating the experiments in a much narrower tank would make the start time clearer, because lateral variations in thickness would be sufficiently small that the area enclosed by the shelf in a photograph would be a good indicator of its volume. 

The application of our model to Newtonian fluids of constant viscosity ($n \equiv 1$) has previously been demonstrated successfully using a Hele-Shaw cell \cite{Pegler_2013} and using a setup rather similar to the one we use \cite{Kowal_2016}. This lends confidence to our results for fluids with a variable viscosity of the form given in Equation \ref{Viscosity_law}.

We observed that all experiments without a sloped bed converged over a length scale that is about $10L$. This suggests that the length of the flat portion of the shelf is only a tenth of the whole shelf, at the time when further convergence towards our similarity solution is no longer discernible in our data. It is possible to estimate $L$ from a photograph $-$ the gradient transitions from 0 to that for the similarity solution over a length scale which is roughly the same as our theory predicts (see \text{Figure} \ref{Real_Profile} and note that the shelf is 90 cm long). As discussed in Section 4.2, convergence should (and does) take longer in experiments with a sloped bed installed (when considering the length of the shelf only).

\begin{figure}[ht]
			\includegraphics [width = 16.5cm] {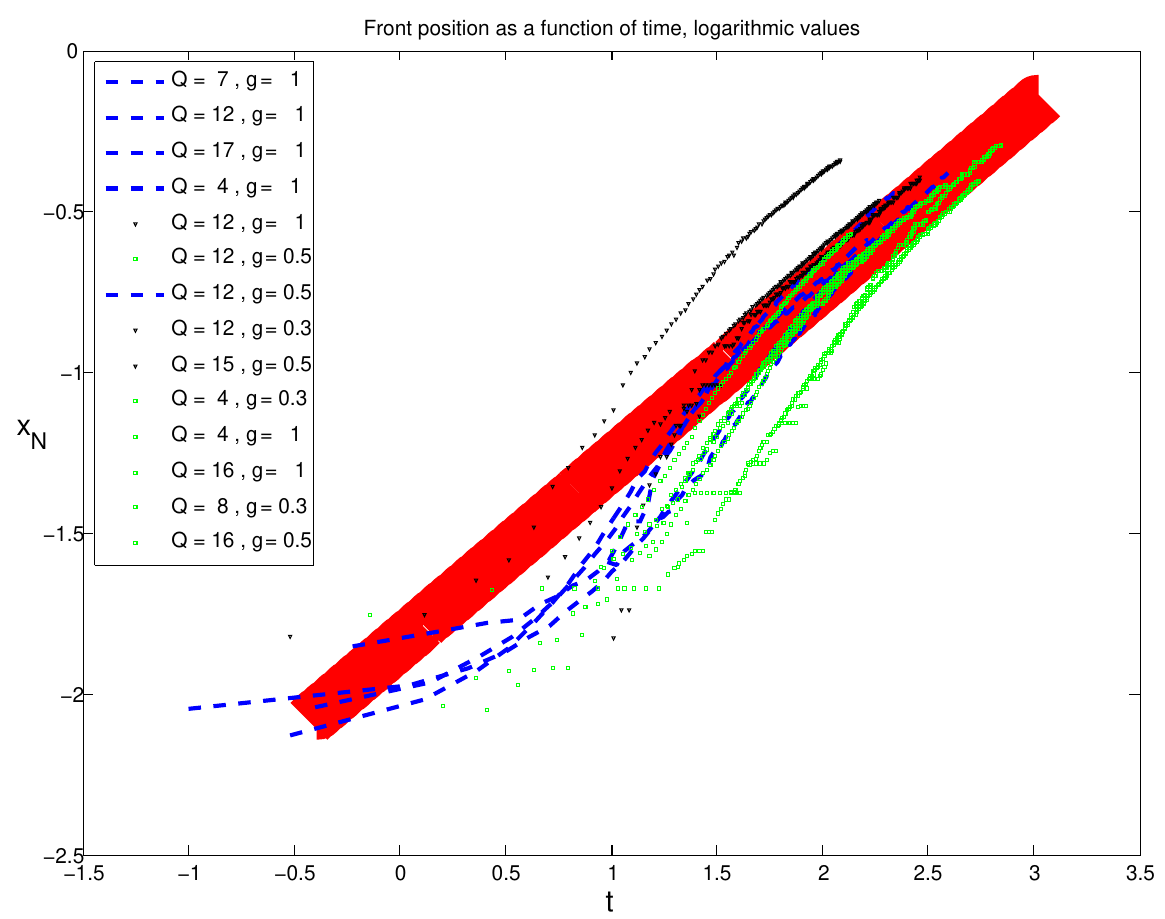}
	\caption{Rescaled front positions as a function of time are shown here on logarithmic axes for our experiments with xanthan at a 0.5\% mass concentration. Blue dashes are from experiments without a sloped bed, while green dots are from ones which had a sloped bed installed. All data is rescaled according to how we expect changes in $Q$ and $g'$ to affect the shelf. The thick red line is our asymptotic (late-time) theory (Equation \ref{Front_position}), allowing for some errors. The black curves are from unreliable experiments $-$ we do not use these data when averaging. Note the extended period in all experiments where the gradient is 1, signifying a constant front speed (before sidewalls eventually make it decelerate). This strongly suggests that, without sidewalls, the front would advance at a constant rate.}
	\label{Half_Percent_Data}
\end{figure}

%The lateral variation in thickness is easy to measure from a photograph, using the colour of the shelf as a proxy for thickness (with calibration possible using the thickness at the sidewalls, which can be directly measured). It is possible to check whether these really are as pronounced as has been supposed, although their effect on ${x}_{n}$ is small.

%Include photographs.

%Determine the multiple of L at which the similarity solution starts becoming valid, with error bars. Mention that it is higher for experiments with a sloped bed installed - as expected.

%Figure 14 = second last one
%Determine n and n_0 to see what the dynamical impact is (it's probably about 10%), and feed in the observed variation (roughly 30%, but check) into my model. This looks roughly right. But, it can't work for 1% xanthan, probably, as the variation is too much.

\newpage
\clearpage
\subsection{Thickness Profile}
\label{Thickness_Profile}

\begin{figure}[ht]
	\centering
		\includegraphics [width = 16.5cm] {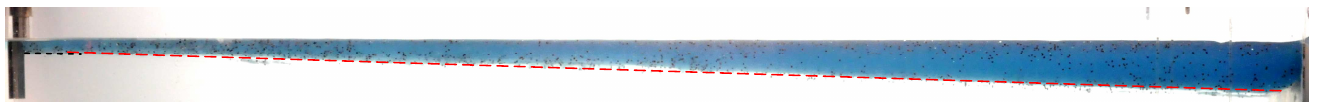}
	\caption{Side view of a sidewall contact experiment (WL1). The dashed red line is drawn to fit the initial gradient of the lower surface of the xanthan. Notice how the shelf is slightly thinner than the red line predicts (i.e. $\left|H'\right|$ rises slightly, as expected). Near the front, the shelf becomes flat. This is because, when this region was near the weir, the theory for ice tongues applied as there was very little sidewall contact. The region should therefore be nearly flat (or parallel to the black line). Notice that this region is a few cm long (for scale, the shelf is 90 cm long) $-$ Equation \ref{L_sidewall} gives $\ssim 3$ cm. At later times, this region then got pushed along by the self-similar (triangular) region of the profile. Things being `pushed along' in this manner is a characteristic feature of our model, because $\frac{\partial u}{\partial x}$ is very small.}
	\label{Real_Profile}
\end{figure}

To see if the profile of the shelf was similar to our expectations (Figure \ref{Similarity_Solution}), we looked carefully at a photograph from near the end of one of our experiments (WL1). This is shown in Figure \ref{Real_Profile}. Not only is the profile very nearly triangular, there is also clear evidence that $\left| H' \right|$ increases along the channel.

\begin{figure}[ht]
	\centering
		\includegraphics [width = 16.5cm] {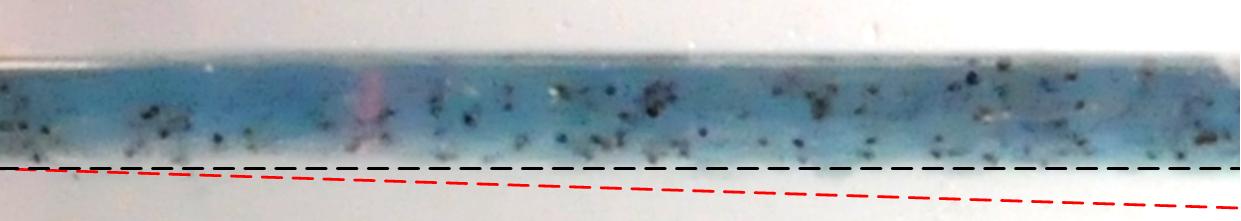}
	\caption{The region near the front (left) in the previous figure is enlarged here. Note that the black line is horizontal, while the red line has the same slope as in the previous figure. As expected, the real profile is not perfectly triangular, with the region shown here in fact being flat. However, this region eventually becomes an insignificant part of the entire shelf, as it doesn't grow. The small reddish region towards the left is a reflection from the laser used to maintain the sea level.}
	\label{Region_Near_Front}
\end{figure}

Near the front, we expect the shelf to be almost flat (Equation \ref{L_sidewall}). Thus, we enlarged this part of Figure \ref{Real_Profile}. This section of the shelf is indeed very close to flat and is certainly much flatter than the rest of the shelf (Figure \ref{Region_Near_Front}).

%Can this page be in landscape?

\clearpage
\newpage
\subsection{Particle Imaging Velocimetry (PIV)}
\label{PIV}

\begin{figure}[ht]
			\includegraphics [width = 7.5cm] {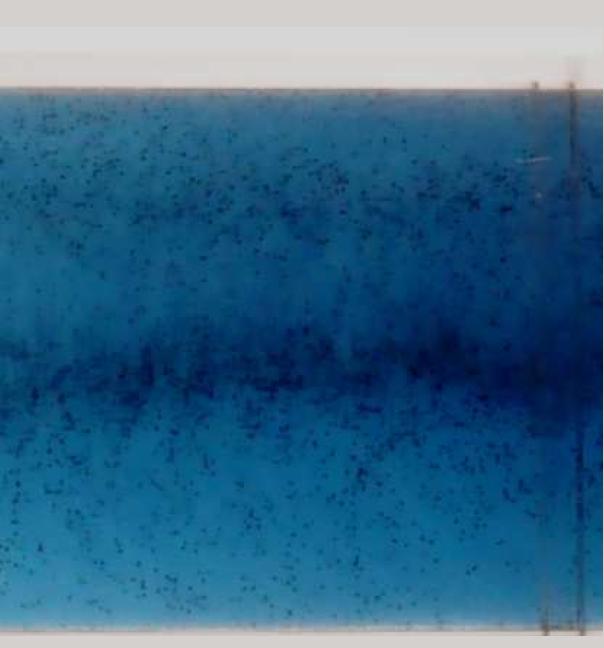}
				\includegraphics [width = 9cm] {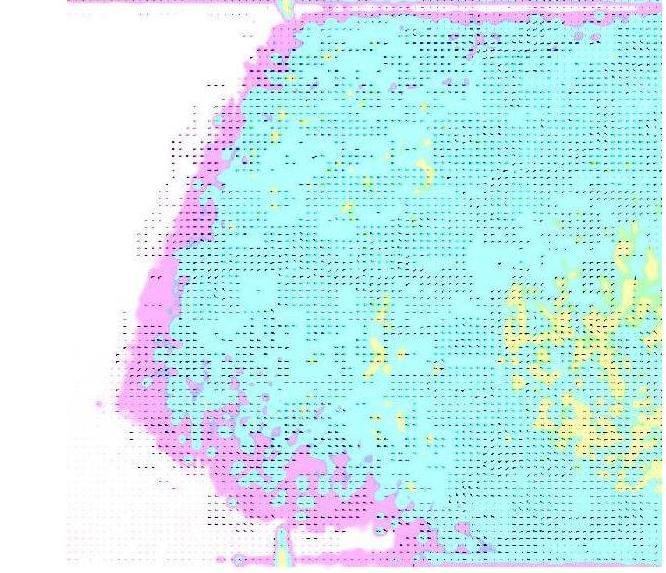}
	\caption{\emph{Left:} The shelf as it appears in a normal camera, viewed from underneath. Note that the central region is a little thicker, and also has more seeds than other regions. Also note that there was no sheet. \emph{Right:} Arrows drawn in by DigiFlow as a result of comparing particle positions between frames. The background indicates the thickness of the shelf (it is in false colour). Different parts of the shelf are shown in the two photographs.}
	\label{Seeds_In_Xanthan}
\end{figure}

%\begin{figure}[ht]
%	\centering
		
%	\caption{The arrows were drawn in by DigiFlow as a result of comparing particle positions between frames. The background indicates the thickness of the shelf (it is in false colour).}
%	\label{PIV_Arrows}
%\end{figure}

%Please include a picture from the first experiment on 17 August 2012 (i.e. at 120 rpm).

We attempted to determine directly whether the velocity profile across the channel in sidewall contact experiments was in agreement with our theoretical model. In order to do this, we put in a large number of poppy seeds into the xanthan (about 0.5\% by mass). These were almost neutrally buoyant, so vertical motion was negligible during the course of our experiments. We used a high-sensitivity black and white camera with a resolution of $1024 \times 1024$ pixels to capture photographs 15 times per second. Later, we used DigiFlow software to analyse these. 

%Thanks to Stuart for inventing it. If you use data from experiments with glass beads, just say this. I haven't used such data (but the PIV on the sheet was done this way).

PIV is a relatively new and difficult technique, so we got a relatively large amount of scatter. We therefore averaged over a 20 second period near the end of an experiment which lasted about 200 seconds. The thickness gradient in the shelf goes down as ${{t}^{-\frac{1}{2n+1}}}$, so we judged that it would hardly change over a 20 second period (and so velocities should hardly change). We also averaged over a 10 cm region of the shelf just beyond the point of maximum thickness. Over this region, the change in $H'$ was minimal (the shelf was $\ssim 8$ times longer than this and is in any case almost perfectly triangular) so $u$ should hardly vary within it.

\begin{figure}[ht]
	\centering
		\includegraphics [width = 16.5cm] {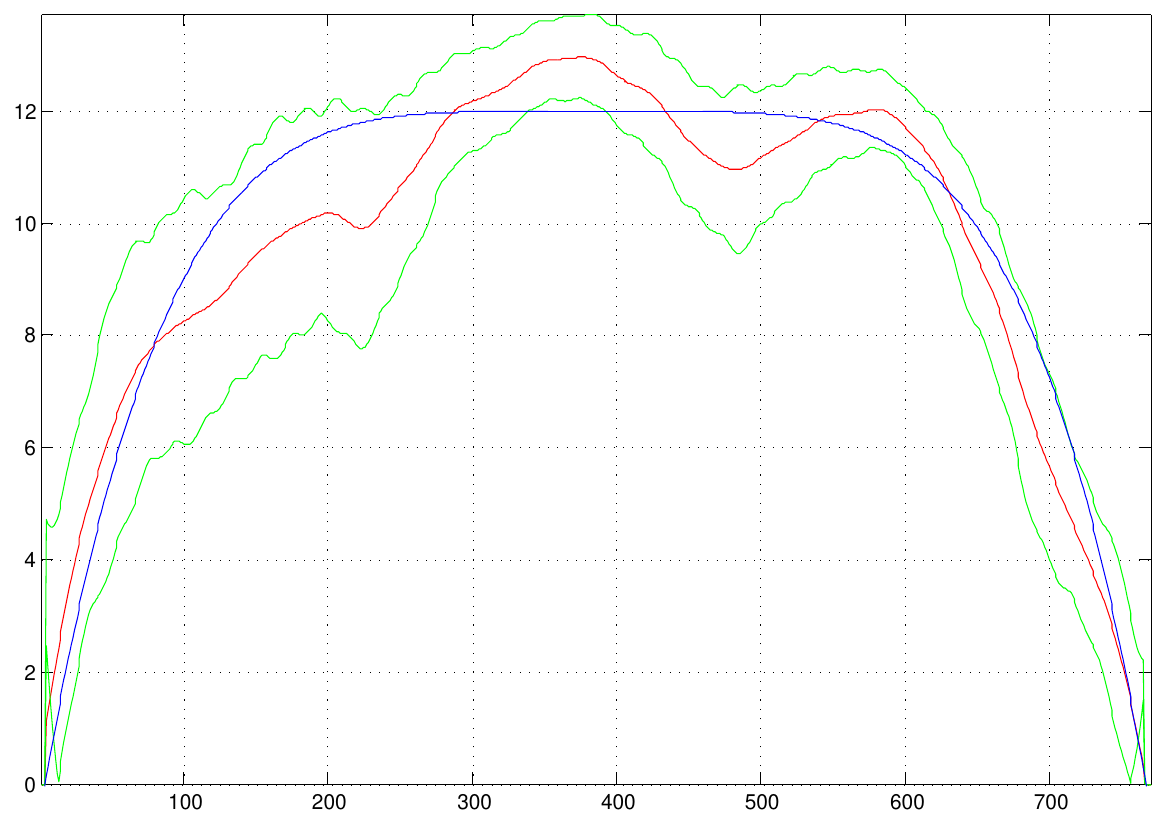}
	\caption{The results of one of our attempts to determine the profile across the channel of the velocity along it, $u \left( y \right)$. The smooth blue line is for theory (Equation \ref{Equation_33}). Observations must lie between the jagged green curves (at $1\sigma$), with raw data along the central red curve.}
	\label{Velocity_Profile}
\end{figure}

The theoretical curve drawn in \text{Figure} \ref{Velocity_Profile} is based on a maximum speed consistent with the PIV data. Because the fluid is clearly satisfying the no-slip boundary condition, the walls of the tank are clearly visible, 763 pixels apart (this corresponds to 15 cm). For the maximum speed, we assume an error of at most 0.5 pixels/second, with a mean value of 12.5. Thus, the maximum velocity is
	\[0.246\pm 0.010\text{ cm/s}\]

The speed of the front at a time concurrent with these observations is found to be 
\[0.27\pm 0.01\text{ cm/s}\]

Errors are because the front decelerates during the 20s we averaged over. The theoretical change in $H'$ between the front and rear of the profile is 2.8\%, corresponding to an increase in $u$ of 11.1\% between the rear and the front (Table \ref{Theoretical_profile_calculations}). This corresponds to a difference in $u$ of 0.027 cm/s, entirely consistent with our observational estimate of $0.024 \pm 0.014$ cm/s.

%Update values if required. Also, check them again.

%Axis labels.

%Please look at the photos from PIV and see where the field of view is, and which temporal part of the experiment it is (use early times when the front was still in view as a guide - also note that the camera was focused just beyond all the grooves, forwards).

We conclude that both lateral and longitudinal variations in $u$ are accurately predicted by our theory. 

\newpage
\clearpage

We also determined vertical velocity profiles in the sheet, to check whether a viscous gravity current model was accurate. We did this in an experiment with sidewall contact, to increase the thickness of the sheet and get more accurate measurements. Otherwise, only a very small number of particles fit in vertically, despite us using particles sufficiently small that a camera right next to the sheet could only just resolve them. 

We expected the viscous gravity current theory to be accurate only for locations sufficiently far downstream of the weir (Equation \ref{L_weir}). Therefore, we were expecting to see a discrepancy between theory and observations sufficiently close to the weir.

%Can include calculations where the pushing force term dominates and there is NO hydrostatic pressure gradient. Should find (with du/dx = alpha du/dz near the base) that du/dz is constant, suggesting a (nearly) triangular velocity profile, which flattens out near the free upper surface of the xanthan. But, this calculation is not really necessary.

We conclude that the viscous gravity current theory is an accurate description of the situation sufficiently far from the weir, but breaks down too close to the weir. This gives us confidence that we have a reasonable understanding of the sheet, suggesting the grounding line may also have been understood.

\newpage
\clearpage
\subsection{No Sidewalls}

We also conducted a series of experiments in which the shelf was not in contact with the sidewalls of the tank, at least for a while. The apparatus still looked the same, except for the weir. This now had a groove cut in the central 5 cm, reduced to 3 cm for the last 2 experiments. The groove was just above sea level. For some of the experiments, we installed the sloped bed in the same way as before (i.e. totally submerged and at $\ssim 10^\circ$).

The front appeared to go at a constant velocity, except at very early times (when all the flux was not yet entering the shelf). This is consistent with our theory, as the experiments ran for much less time than that required for convergence to a self-similar mode of propagation. Consequently, the front propagated at constant speed. Combined with what appeared to be a constant width to the shelf, we suppose that the grounding line had reached a dynamic equilibrium.

Due to severe technical difficulties, 6 experiments are not shown because the shelf rapidly hit a wall of our tank. Readers attempting to repeat our experiments should note that the inlet and seawater extraction pipes should be (extremely close to) vertical and in the middle of the tank and the whole tank (and any sloped bed inside it) should be level to the horizontal to within about 3 arc-minutes. The sloped bed, which needs to rest against the weir, also needs to have been manufactured to the correct working angle (within a few degrees).

In what follows, we assume that the constant velocity of the shelf can be used to determine its thickness at the grounding line, given also its width and the supplied flux. One minor complication in measuring the width is that we used the bottom view of the shelf (obviously), whereas the length of the shelf came from the side view. This is because a thin shelf is hard to see from underneath, but it still provides 15 cm of optical depth when viewed side on. However, the optical paths are different in the two cases due to an extra reflection. This means the same physical distance appears as a greater number of pixels in the camera focal plane for the side view. 

Errors result from lateral variations in thickness, which make $d$ hard to determine. These variations were enhanced by the tendency of the fluid to spread sideways without any lateral confinement. We used the colour of the shelf in the bottom view to determine which regions were thin. These regions have been excluded from our measurements of $d$.

Also, the grounding line is not at constant thickness laterally (i.e. it isn't a `line'). This led to a systematic difference between theory and measurements. The reason is that the thick central regions of the sheet at the grounding line, which our theory addresses (because these regions make up most of the sheet), were thicker than the shelf far downstream. Apparently, after the grounding line, the shelf in these regions thins with distance until it becomes roughly the same thickness as the \emph{thinnest} parts of the sheet at the grounding line. Thus, the velocity measurements were essentially indicating how thick the thinnest regions of the grounding line were. This causes the theoretical grounding line thickness (for the thick central regions) to exceed the `measured' thickness (for the thin regions near the edge). This is an interesting phenomenon, and again a case of lateral structure in the flow outside direct consideration in our model having an influence on the shelf. As before, the effect is more pronounced with a more viscous fluid (more concentrated xanthan).

\begin{figure}[ht]
	\centering
		\includegraphics [width = 16cm] {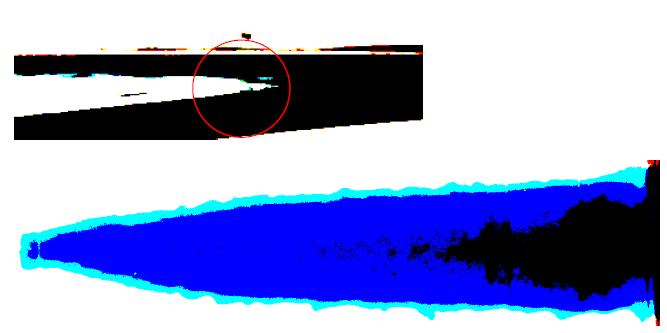}
	\caption{This is a contrast-enhanced photograph of an experiment. The thicker regions appear darker in the bottom view. Note that downstream (left) of the thickest region of the grounding line (in the middle), the shelf thins until it is approximately the same thickness as the thinnest regions of the grounding line. The thinning is evident from the side as well (circled region in top panel).}
	\label{Contrast_Enhanced_Ice_Tongue}
\end{figure}

Our experiments indicated that there was a large amount of lateral spreading in the sheet, mostly very close to the weir (even though the sloped bed was entirely underwater, which we believe reduces the spreading). However, it appeared that there was no noticeable lateral spreading in the shelf. In fact, the spreading appeared to have occurred well upstream of the grounding line in all of our experiments. Combined with the constant speed of the front of the shelf, this strongly suggests a constant thickness.

We attempted to determine directly the thickness of the shelf at the grounding line. The resolution on this was relatively poor, because the thickness is only a few mm in most cases. Thus, it appeared in our photographs as about 20 pixels. Reflections from the ocean and a small amount of parallax made it extremely difficult to perform this sort of measurement (as a look at the photographs will show). The much lower accuracy prevented us from getting a reliable indication of whether our theory is correct using such measurements. However, they did indicate that direct measurements of ${H}_{G}$ made in this way are consistent with the results of measuring the downstream shelf velocity and width, which we could determine more accurately and used for testing our theory instead.

Experiments were also conducted without a sloped bed. These indicated negligible lateral spreading, suggesting that ice tongues fed by sheets on steep slopes are unlikely to be much wider than the sheet.

%\begin{table}[ht]
%	\centering
%		\begin{tabular}{ccccccccc}
%\hline
%Expt. & Flux & Slope & $\rho_w$ & ${H}_{G}$ & Error & ${H}_{G}$ & Error & Prediction & Ratio\\
%& & of bed & & (mm) & & (from & & &\\ 
%& & (degrees) & & & & front) & & &\\
%\hline
%H1 & 3.22 & 6.5 & 1100 & 0.7 & 7.4 & 0.3 & 6.4 & 1.15\\
%H2 & 10.05 & 8.4 & 1100 & 2.4 & 9.1 & 0.8 & 7.5 & 1.20\\
%BH3 & 7.90 & 9.91 & 1100& 9.4 & 1.2 & 10.6 & 1.0 & 13.2 & 0.80\\
%BL4 & 3.71 & 8.25 & 1100 & 7.2 & 1.2 & 6.8 & 0.5 & 6.7 & 1.02\\
%BL5 & 3.88 & 8.59 & 1175 & 4.3 & 1.0 & 6.3 & 0.5 & 6.4 & 0.98\\
%\end{tabular}
%\caption{Results obtained from experiments without lateral friction. The ratio is observations/prediction. Note that BL4 and BL5 had different $g'$. Also note that we have only a very basic understanding of experiments without a sloped bed (the first two listed above).}
%\end{table}

%The data is as in the photo, definitely.

\newpage

We also attempted to understand experiments without a sloped bed. Based on high-resolution photographs, we believe that the angle of the upper surface of the xanthan behind the weir is always close to $30^\circ$. Combined with the no penetration condition, we can obtain a velocity profile analogous to those obtained previously for the sheet. The thickness required to drive flux $Q$ through the weir is then assumed to equal the downstream shelf thickness.

However, it is almost completely certain that other phenomena are critical to understanding the weir. In particular, surface tension has implicitly been included in our model (in terms of fixing the angle of the free surface) but not explicitly in the force balance. This means that we can not expect the predictions based on this theory to be very accurate. On the other hand, it does give the right order of magnitude. Among other things, future work will need to predict an angle of the upper surface close to observed values.

For experiments with a sloped bed, we find good agreement with 0.5\% xanthan. At 1\%, although there is only one experiment, it is again highly probable that lateral variations in thickness are more pronounced. As these are outside our model, they will lead to a reduction in the accuracy of our predictions.

The prediction that $g'$ does not significantly affect the grounding line thickness appears to be borne out by a comparison of experiments BL4 and BL5, the last of which nearly doubled $g'$ relative to the others. However, both theory and experiments show a slight reduction in grounding line thickness as a result of making the xanthan nearly twice as buoyant.

Another thing we note is that the shelf buckled in one of our experiments. This led to the front alternately hitting one wall and then the other. The reason for this is unclear, but it appears to be due to internal elasticity of the xanthan. We expect that this is unlikely to occur in ice. The buckling had a small effect on the speed of the front, but much smaller than the error in $d$, so we do not discuss it further.

We now show that the effect of non-hydrostatic forces from the weir was dissipated against basal friction before the location of the grounding line. Xanthan is assumed to overflow the weir by an amount H. We assume that forces here are only hydrostatic, but that this gets converted to a non-hydrostatic force on the sheet due to the highly artificial geometry in the situation. Genuine non-hydrostatic forces at the weir may be calculated on the basis of $h'$ always being approximately $30^\circ$. Such forces appear to be negligible in all of our experiments, compared with hydrostatic forces.

The vertically integrated hydrostatic pressure is $\frac{1}{2}\rho g{{H}^{2}}$. The basal friction per unit length is ${{\eta }_{o}}{{\left( \frac{1}{2}\frac{\partial u}{\partial z} \right)}^{\frac{1}{n}-1}}\frac{\partial u}{\partial z}$. We assume that
	\begin{eqnarray}
	\frac{\partial u}{\partial z}&\approx &\text{ }\frac{2 \overline{u}}{H}\text{  } \text{    (vertical average indicated)} \\
&\approx&\frac{2Q}{{{H}^{2}}d}
\end{eqnarray}

This is an underestimate because the upper surface of the sheet is free so the base must have more than average shear. However, this will not affect the final result much if $n$ is large (because $n^\frac{1}{n} \to 1$ as $n \to \infty$).

We obtain that the non-hydrostatic force exerted by the supply mechanism will be dissipated over a length scale $L$, where
\begin{eqnarray}
	L ~=~ \frac{\rho g{{H}^{2+\frac{2}{n}}}}{{{4\eta }_{o}}}{{\left( \frac{d}{Q} \right)}^{\tfrac{1}{n}}}
	\label{L_weir}
\end{eqnarray}
	
For $H$ around 5 mm at the weir (as suggested by photographs) in a 0.5\% xanthan experiment, we get that $L<2$ cm. This corresponds to a grounding line that needs to be at least 4.5 mm thick (including an allowance for the sea level being 1 mm higher than the top of the slope). We obtain similar conclusions for experiments at 1\%, allowing the thickness at the weir to rise to as much as 9 mm to account for the greater viscosity. However, the greater viscosity of the fluid also makes it easier for basal friction to dissipate the force exerted at the weir. Thus, noting that $L$ has been overestimated, we conclude that our grounding lines can not have been significantly affected by the force exerted at the weir. Therefore, a simple viscous gravity current model for the sheet should suffice for determining $H_G$.

We also believe it unlikely that realistic natural situations will allow for such unusual configurations as we had in our experiments, especially anything resembling our weir. Therefore, ice sheets can likely be understood solely in terms of hydrostatic pressure gradients balancing basal friction. However, this should be confirmed based on real viscosity parameters and topography.

\begin{figure}[ht]
			\includegraphics [width = 11.5cm] {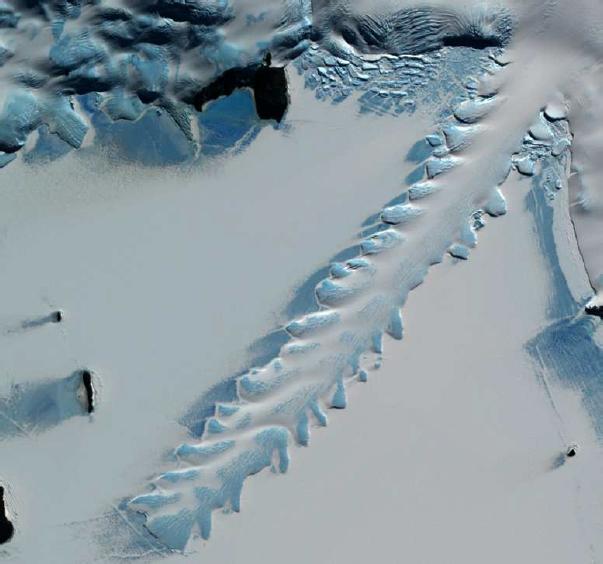}
			\includegraphics [width = 4.2cm] {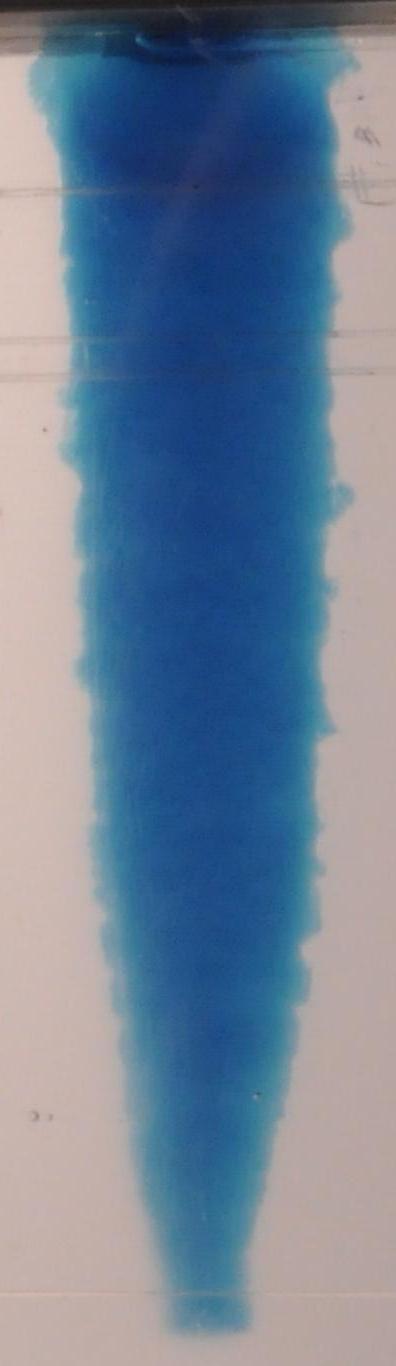}
	\caption{The Erebus ice tongue (\emph{left}), showing a similar edge to our laboratory model for such systems (\emph{right}). We believe that we mimicked a seasonal variation in the entry flux with the oscillatory action of our peristaltic pump, leading to the similar appearance. Note that our model shelf is much wider than the groove in the weir. This suggests that the shelf determines its own width, this being affected by $Q$ and likely also by other parameters.}
	\label{Comparison_with_Erebus}
\end{figure}

A final note concerns the unusual corrugated pattern of the edge of the flow in our experiments. This is known to occur in nature (Figure \ref{Comparison_with_Erebus}). Our understanding is that the oscillatory action of our peristaltic pump causes variations in flux which lead to variations in the thickness of the shelf at the grounding line, presumably because a higher flux causes the sheet to spread laterally by an increased amount. As the rest of the shelf essentially moves as a rigid body, the pattern remains permanently imprinted upon the shelf. This may also explain why the front of the shelf tapers $-$ it crossed the weir when the flux overflowing it was still rising towards its final value.

\begin{figure}[ht]
	\centering
		\includegraphics [width = 11.5cm] {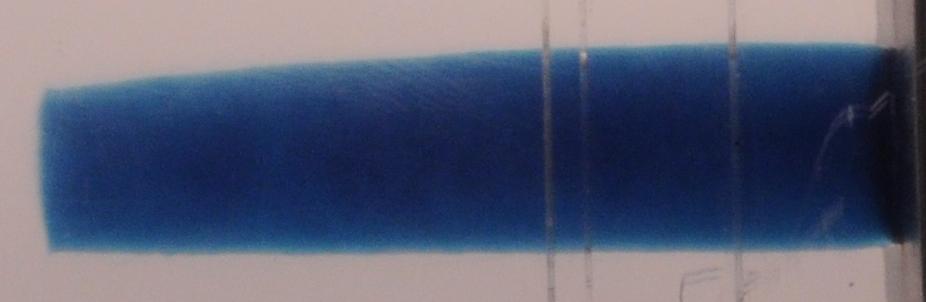}
	\caption{In this experiment, there was no sloped bed. The corrugations are now absent, although the same pump was used for all of our experiments. The shelf is now only as wide as the groove in the weir.}
	\label{No_Sloped_Bed}
\end{figure}

In natural situations, the effect can be due to seasonal or other changes in the flux entering the ice shelf. We note that the effect did not arise in those experiments in which we did not install a sloped bed (Figure \ref{No_Sloped_Bed}). Thus, the effect is likely reduced if the terrain is steeper close to the grounding line. This affects the area in contact with the ocean, which may have important consequences.

\newpage
\clearpage
\section{Towards The Natural System}
\label{Towards_nature}

\subsection{Including Flux Non-Conservation}

Real ice shelves do not grow forever. This is because ice is lost from them. There are two major ways in which this occurs: iceberg formation and basal melting. We will neglect sublimation at the top compared to these processes.

Melting on the underside of an ice shelf is an important process. We assume that it proceeds at a rate proportional to the surface area in contact with the ocean. Stability of this surface is assumed. As the ice shelf is nearly flat, we assume the area remains the same if projected vertically ($\cos H' \approx 1$).

Calving of icebergs at the front of an ice shelf is still poorly understood. We treat it in a similar way to melting on its underside. Thus, the rate at which volume is lost is also proportional to the surface area of the front of the ice shelf.

The exact mechanism by which this occurs will turn out to be important, in particular whether a higher temperature is likely to increase it by a similar amount to the rate of melting on the underside. For the moment, we assume that ocean water in contact with the front of an ice shelf slowly melts it. This may happen most efficiently near the ocean surface. As the ice is melted away, a large overhang will be left. This will eventually collapse. If the melting primarily occurs in a narrow layer near sea level, then there will also be an `underhang'. Due to buoyancy forces, this will eventually break off as well.

The most widely believed alternative to this mechanism is the action of waves at the front of an ice shelf. It seems likely that this mechanism will also lead to loss of volume being directly proportional to surface area. However, it is less sensitive to temperature.

Our basic model therefore has two additional parameters governing these mechanisms of ice loss. Using $F$ to denote the volume rate of loss of ice from the shelf (and $A$ for area), we define
\begin{eqnarray}
	c &\equiv& \frac{\partial F}{\partial A} \text{    (basal melting)} \\
	f &\equiv& \frac{\partial F}{\partial A} \text{  (calving of icebergs)}
\end{eqnarray}

\newpage
\clearpage
\subsection{Ice Tongues}

Ice tongues are nearly flat due to the absence of substantial drag on any part of the shelf. Although there is theoretically a small amount of thinning between the grounding line and the front of the shelf (see Equation \ref{Ice_tongue_profile}), we will assume here that ice tongues are perfectly flat owing to the very large value of $L$.

The presence of basal melting will not substantially affect this, although it will require $u$ to decrease with $x$. This will lead to some longitudinal stress, but we assume that this does not much affect the force balance. It is also evident that a region of the shelf which is thinner than regions upstream of it will eventually thicken due to the increased flux entering this region. Thus, the ice shelf should be stable.

The mass balance for an ice shelf will approximately be given by
\begin{eqnarray}
	Q ~=~ d(cx_n + H_Gf)
\end{eqnarray}

We will assume that $d$ will not change much. Most likely, it is set by topography close to the grounding line. Thus, an increase in $c$ or $f$ will lead to a reduction in $x_n$. However, assuming that conditions in the interior of the continent have not changed much, $Q$ will still be the same and so $H_G$ will remain unaltered (Equation \ref{H}).

This means that a long ice tongue is not in imminent danger of collapse: it first needs to shorten. If such a collapse were to occur anyway, the force balance at the grounding line would be unaffected and so there would be no reason for the flux entering the ocean to change. Most likely, the ice tongue would re-establish itself.

Under some circumstances, $f$ may become sufficiently high that the ice tongue ultimately has its length reduced to 0. In this case, a small further rise in $f$ will cause the grounding line to retreat. The force balance will now be affected. The resulting reduction in the integrated hydrostatic pressure of seawater will require a reduction in the pushing force (Equation \ref{G}). This means the flux entering the ocean must be reduced. 

However, with a continual supply of ice, it is not possible for the flux entering the ocean to actually be reduced. The ice sheet will attempt to steepen to keep the flux entering the ocean equal to that carried away by icebergs. This will require the upper surface to steepen, achieved by loss of ice near the grounding line. Such a loss is in any case required to maintain the flotation condition, which we believe will still hold at the grounding line.

According to Equation \ref{H}, however, it is not possible for there to be an equilibrium solution at such a reduced grounding line thickness. The basic reason is that, for forces to balance, the upper surface of the sheet must be roughly parallel to the sloped bed. The extra surface gradient now present will not be sustainable in our model. We suppose that ice is also lost further upstream than the grounding line, thereby maintaining the gradient of the upper surface.

This reduces the flux entering the ocean (Equation \ref{Q}). It also reduces the amount of ice lost to iceberg formation. However, it is clear that the latter effect is much smaller than the former. Consequently, the grounding line will be forced to retreat even further.

Without doing detailed calculations (which may well give a different outcome), we speculate that the equilibrium solution is for the grounding line to be above sea level. Then, it is possible for the ice sheet to have zero thickness at the grounding line (buoyancy forces prevent this occurring for a grounding line below sea level). This will mean the pushing force at the front is zero, but some flux enters the ocean if the front is very steep (as will likely occur). An estimate of the timescale for reaching this equilibrium may be obtained by setting $Q = Hf$. Eventually, of course, the flux entering the ocean and that supplied to the sheet will be equal.

%As a rough idea, the ocean is too hot to allow any ice below sea level. The instability we have just described appears to be more severe for a shear-thinning fluid, due to the strong dependence of $Q$ on $H$.

\begin{figure} [ht]
%\centering
  \includegraphics[width = 17cm]{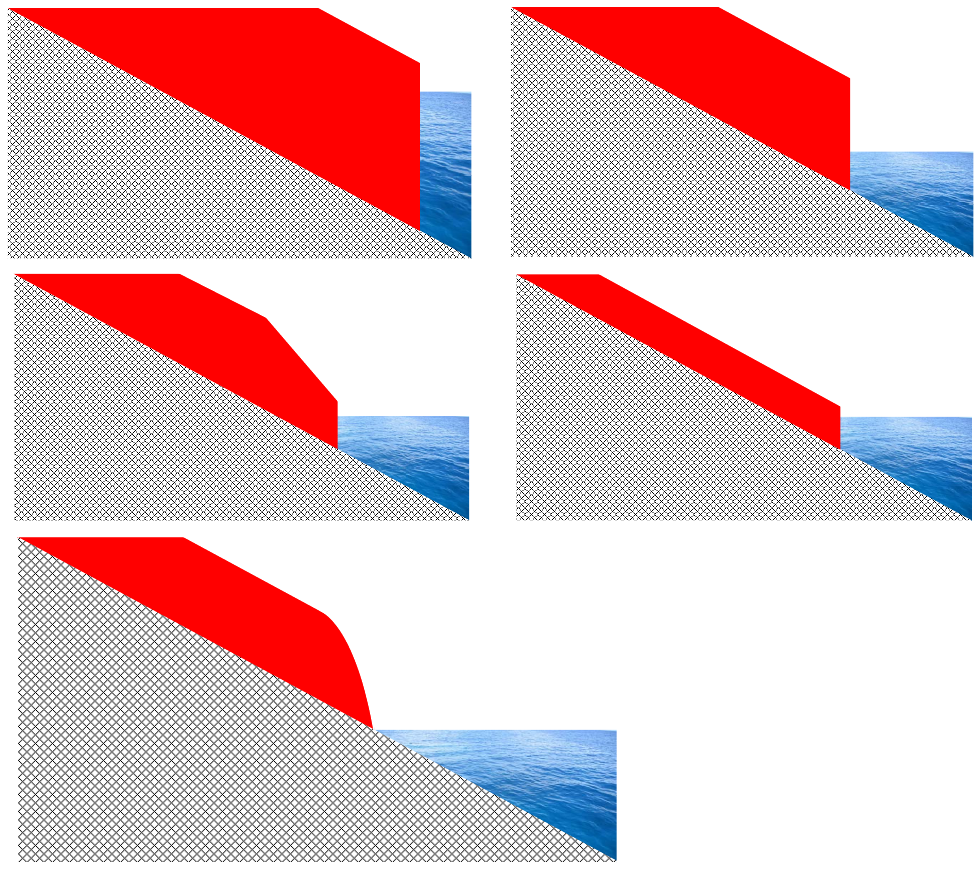}
\caption{The ice sheet is initially in equilibrium, with the flux entirely carried off by icebergs (so no shelf). The sheet runs nearly parallel to the sloped bed. A small further increase in iceberg formation $f$ causes recession of the grounding line (panel 2). This also causes the front of the sheet to steepen (panel 3). The steeper gradient leads to a higher flux, removing ice further upstream and reverting the sheet to the previous surface slope (panel 4). In this situation, the forces may be balanced but icebergs are carrying far too much ice away because the supplied flux is much more sensitive to the thickness than the loss due to iceberg formation (Equation \ref{Q}). Thus, the grounding line recedes further. We suppose that it ends up very close to sea level.}
%\label{State_1}
\end{figure}

\newpage
\clearpage
\subsection{Laterally Confined Ice Shelves}

A laterally confined ice shelf requires a gradient in its thickness in order to flow. For the moment, we neglect the formation of icebergs and assume that sidewalls dominate the force balance.

Setting $x = 0$ at the front of the ice shelf and using $q = 2dcx$ (assuming zero thickness at the front), we see that $H{H'}^{n} \propto x$. Separating the variables, this implies that the shelf will be perfectly triangular. A quick look at any of our experiments suggests this to be perfectly reasonable (Figure \ref{Real_Profile}), although none of them involved loss of fluid in this way.

In this model, the entry flux fully determines the length of the shelf.
\begin{eqnarray}
	x_n &=& \frac{Q}{2cd}\\
	H' &=& \frac{2cdH_0}{Q}
	\label{Equation_68}
\end{eqnarray}

Combining this result with Equation \ref{Shelf_flux}, we see that
\begin{eqnarray}
	{{H}_{0}} ~=~ Q{{c}^{-\frac{n}{n+1}}}{{\left( \frac{n+2}{4} \right)}^{\frac{1}{n+1}}}{{\left( \frac{{{\eta }_{o}}}{\rho g'} \right)}^{\frac{1}{n+1}}}{{d}^{-2}}
	\label{Equation_69}
\end{eqnarray}

As with ice tongues, an increase in $c$ will force a reduction in $x_n$. This time, however, $H_0$ will also decrease. This will lead to grounding line retreat. To maintain the same entry flux, $H'$ will be forced to increase.

The solution we have found is self-consistent as no icebergs will form at the front. We now check whether this solution is stable. Suppose that a small region of shelf near the front broke off. We keep the origin of our co-ordinate system where the front previously was. The condition for stability is that more flux crosses the front than can be carried away by icebergs, thereby leading to a longer shelf and restoration of the lost region. This means that
\begin{eqnarray}
	2dcx ~>~ 2dfH'x \label{Shelf_Stability}
\end{eqnarray}

%If other mechanisms are responsible, then $f$ will perhaps rise slowly or not at all despite large rises in $c$. The distinction will turn out to be critical. (New paragraph) If $f \propto c$, then 

If melting of ice is primarily responsible for the formation of icebergs at the front, then it is reasonable to have $c \propto f$. In this case, increases in these parameters will force an increase in $H'$ and make Equation \ref{Shelf_Stability} harder to satisfy. The shelf will eventually become unstable. Unlike in the case of ice tongues, a large laterally confined ice shelf can suddenly become unstable and rapidly disintegrate. This will cause a drastic alteration in the force balance at the grounding line. The buttressing effect of the sidewalls upon the system is now gone. Consequently, the system will behave more like an ice tongue. This would most likely mean the new equilibrium thickness will be given by Equation \ref{H}.

We suppose that, before the sheet has quite reached equilibrium, it will attempt to balance the forces across the grounding line. This will be achieved by having the ice run parallel to the bed very close to this point. This allows for a rough estimate of the flux entering the ocean, using Equation \ref{Q} and setting $h' = \alpha$. One can of course calculate more precisely what flux is required for a particular value of $H$ to be the equilibrium grounding line thickness. As this flux will undoubtedly greatly exceed the flux supplied into the ice sheet, the grounding line will retreat. Our equations can thus be used to get a rough estimate of how long it will take to reach the new equilibrium configuration.

The grounding line will most likely feed an ice tongue. We believe it unlikely that the rate of iceberg formation will be sufficiently high to prevent this occurring. Also, it appears unlikely that sidewall contact will be properly re-established. However, if $f$ was high enough, an ice tongue would not form at all. The equilibrium configuration would then not be as just discussed. Most likely, the grounding line would retreat all the way to sea level, with the retreat rate governed by the efficiency of iceberg formation ($Q = 2Hdf$).

Our reasoning is based on assuming that $H' \propto c^{\frac{1}{n + 1}}$ (Equations \ref{Equation_68} and \ref{Equation_69}) and that $f \propto c$. However, if $f$ and $c$ are not linked in this way, it is possible that $f$ rises slowly or not at all even as $c$ increases substantially. One possibility is that basal melting determines $c$ while the mechanical stresses induced by waves primarily determine $f$, such that a temperature rise increases the former but does not much affect the latter. In this case, Equation \ref{Shelf_Stability} may well continue to hold as $c$ rises. Thus, it is important to obtain a better understanding of the processes occurring at the front of ice shelves and on their undersides. We recommend giving particular attention to the question of whether they respond differently to changes in environmental conditions.

\newpage
\clearpage
\section{Conclusions}

Recent breakthroughs in understanding the force balance in a simplified laboratory model of a marine ice sheet \cite{Pegler_2013} have now been significantly extended. The theory we developed is valid for the case of a shear-thinning power law fluid (Equation \ref{Viscosity_law}) with arbitrary index $n$, not just for a Newtonian fluid with $n \equiv 1$. Previous experiments suggest that a fluid of this type with $n \approx 3$ provides a good description of the behaviour of ice in glaciers \cite{Glen_1955}. Laboratory experiments confirm that our theory is valid to within the very tight experimental tolerances we achieved. These experiments revealed additional aspects of the laboratory model that are still not fully understood, such as buckling and lateral thickness variations. We consider it unlikely that buckling could happen in natural ice shelves, though lateral thickness variations may play an important role (Figure \ref{Comparison_with_Erebus}). We believe such variations are caused by seasons in natural systems and that we unintentionally mimicked such effects with the action of our peristaltic pump. Our experiments suggest conditions in which this effect is greatly reduced (Figure \ref{No_Sloped_Bed}).

Our theory for ice tongues is based on a straightforward generalisation of the work of Robison \cite{Robison_2010}. The equilibrium grounding line thickness is found in a similar way, but the additional complexity means there is no true analytic solution. We do, however, find an approximate analytic solution. The basis for our approximation scheme is that the upper surface of the sheet should be parallel to the lower surface (the bed). This is not always true, so we derive conditions for it to be a reasonable approximation (Equation \ref{Equation_59}). For ice in water, the bed should have a slope of $\ssim 9^\circ$.

For shelves that are laterally confined, our theory is based on the assumption that the fluid undergoes generalised Poiseuille flow (Equation \ref{Equation_33}). This is true for sufficiently long shelves, assuming they do not lose contact with the sidewalls. We obtain constraints on what length is required (Equation \ref{L_sidewall}), with observations indicating good agreement with our predictions and especially the correct dependence on parameters like the entry flux (Figure \ref{Half_Percent_Data}). Our predictions for a Newtonian fluid have been confirmed by other workers using a very narrow Hele-Shaw cell to guarantee sidewall domination \cite{Pegler_2013} and using a setup rather similar to the one we use \cite{Kowal_2016}.

Experiments confirm that the presence of a grounding line does not affect the asymptotic behaviour of laterally confined shelves. Therefore, our prediction for the shelf thickness at its start is equivalent to a prediction of the grounding line position. In this case, an important feature of our solution is that the grounding line advances for ever, although it decelerates.

We combined our understanding of the forces with a basic model for loss of ice from a shelf due to melting on its underside and iceberg formation at its front. This prevents the shelf growing for ever, such that an equilibrium configuration is eventually attained. However, there is an instability peculiar to laterally confined ice shelves, which can suddenly collapse if oceanic conditions change in a particular way. Ice tongues seem more stable, but if conditions alter substantially and prevent one from existing at all, then the associated sheet becomes unstable. This may cause retreat of the grounding line to sea level, although this process may take a long time.

%Future work will need to concentrate on integrating our model for the force balances (and the velocity fields they produce) with a model for other types of mass transfer. In real ice shelves, melting and snowfall play an important role, but especially crucial is the calving of icebergs at their front. This somehow prevents real ice shelves from thickening for ever, meaning it is of fundamental importance to the shelf (just as much as the force balance is).

%The buttressing exerted by the shelf upon the sheet, which supposedly causes significant acceleration of the sheet if it is removed;
It is often thought that removal of the buttressing exerted by the shelf upon the sheet would cause the sheet to accelerate significantly. However, this buttressing actually comes from sidewall contact. If there is no sidewall contact, then the buttressing is equivalent to what would be provided by hydrostatic pressure of water alone, which means this will still be present with no shelf. Only the additional amount due to the shelf being in contact with sidewalls can be removed by melting the shelf, so in ice tongues we do not expect a sudden acceleration if the shelf were to break up. In this case, a reduction in viscosity (e.g. due to global warming) may still cause a significant acceleration of the flow because this depends on $\eta_o$, a very temperature-sensitive quantity for ice close to its melting point \cite{Glen_1955}. However, such an accelerated flow would not be due to collapse of the ice shelf.

The application of our model to marine ice sheets requires assumptions about poorly understood processes like iceberg formation. Detailed understanding of these processes is essential if we are to fully understand events like the collapse of the Larsen B ice shelf. Hopefully, these events can be understood in time to prepare for their consequences and perhaps to alter our actions to make them less likely. We hope the insights presented in this contribution will be helpful in this regard.

\newpage
\clearpage
\section{Acknowledgements}

The authors wish to thank the technicians of the GK Batchelor Laboratory for producing several components of fundamental importance to this project and for suggesting a method to achieve better levelling of the tank with the horizontal. They also thank the referee for their input. IB is supported by a Science and Technology Facilities Council studentship. The algorithms were set up using \textsc{matlab}$^\text{\textregistered}$.

\end{document}